\documentclass[aps,prb,reprint,preprintnumbers,twocolumn,footinbib]{revtex4-1}
\usepackage{graphicx}
\usepackage{bm}
\usepackage{epsfig}
\usepackage{enumerate}
\usepackage{ulem}
\usepackage{color}
\usepackage{mathrsfs}
\usepackage{dcolumn}
\usepackage{setspace}
\usepackage{array}
\usepackage{amsmath}
\allowdisplaybreaks[4]
\usepackage{amssymb}
\usepackage{dsfont}
\usepackage{gensymb}
\usepackage{dcolumn}
\usepackage{multirow}
\usepackage{breakcites}
\usepackage{bibentry,natbib}
\usepackage{booktabs}

\begin{document}
\title{Vacuum-Induced Symmetry Breaking of Chiral Enantiomer \\ Formation in Chemical Reactions}

\author{Yanzhe Ke$^{1,2,\dag}$}
\author{Zhigang Song$^{3,\dag}$}
\author{Qing-Dong Jiang$^{1,4,}$}
\email{qingdong.jiang@sjtu.edu.cn}
\affiliation{{}\\ $^1$ Tsung-Dao Lee Institute and School of Physics and Astronomy, Shanghai Jiao Tong University, Shanghai 200240, China\\
	$^2$ Department of Physics, Hong Kong University of Science and Technology, Clear Water Bay, Hong Kong, China\\
	$^3$ John A. Paulson School of Engineering and Applied Sciences, Harvard University, Cambridge, Massachusetts 02138, United States\\
	$^4$ Shanghai Branch, Hefei National Laboratory, Shanghai 201315, China\\
	$^\dag$These authors contributed equally to this work\\
}
\begin{abstract}
	
	A material with symmetry breaking inside can transmit the symmetry breaking to its vicinity by vacuum electromagnetic fluctuations. Here, we show that vacuum quantum fluctuations proximate to a parity-symmetry-broken material can induce a chirality-dependent spectral shift of chiral molecules, resulting in a chemical reaction process that favors producing one chirality over the other. We calculate concrete examples and evaluate the chirality production rate with experimentally realizable parameters, showing the promise of selecting chirality with symmetry-broken vacuum quantum fluctuations.
	
\end{abstract}
%\preprint{MIT-CTP/xx}
\maketitle

{\it Introduction.---} 
The notion of chirality (or handedness) dates back to the year 1848 when Louis Pasteur noticed two types of crystals, each one a mirror image of the other. Since then, the ubiquitous existence of chirality has been recognized and appreciated in different areas ranging from fundamental physics to chemistry and biology. One of the most promising endeavors is to explain the origin of molecular handedness in nature\cite{hegstrom1990handedness}. 

Chemists often refer to mirror-image molecules as \textit{L} enantiomers and \textit{D} enantiomers, where \textit{L} and \textit{D} conventionally stand for left- and right-handedness.
When chiral molecules are synthesized from achiral building blocks, equal amounts of the \textit{L} and \textit{D} enantiomers are usually produced in the absence of external influences. 
Therefore, chirality selection in chemical reactions remains an important but painstakingly difficult task\cite{avalos1998absolute}. 

The most common approach to selecting chirality in chemistry is by using a chiral catalyst in a sophisticated procedure\cite{maier2001separation,noyori2003asymmetric}. However, it is difficult to identify the appropriate chiral catalyst for a specific chemical reaction. Additional methods involve employing external influences, such as static EM fields or circularly polarized radiation, to discriminate chirality\cite{balavoine1974preparation,power1974circular,huck1996dynamic, salam1997control, shao1997control, salam1998enantiomeric, hoki2002quantum, PhysRevLett.86.1187, ma2006chiral, PhysRevLett.110.213004,PhysRevLett.113.033601,xiaopeng,hollander2017,PhysRevA.105.053711}. The efficiency of these approaches depends on specific circumstances \cite{avalos1998absolute} and typically does not exceed the efficiency of the catalytic approach \cite{inoue1992asymmetric,rikken2000enantioselective}. 
%whose efficiency is (current in debatable\cite) and generally not comparable/unmatchable to the catalytic approach\cite{inoue1992asymmetric,rikken2000enantioselective}.}
%{\color{red}which are generally no more efficient than the chiral catalytic approach\cite{inoue1992asymmetric,rikken2000enantioselective}.} 
Therefore, the quest for a universal and more efficient method to select chirality is of great practical importance. 

Vacuum quantum fluctuations are a promising candidate for selecting chirality in chemical reactions. At first glance, it would appear impossible since a vacuum, by definition, contains nothing and would not affect chemical reactions.
However,  quantum fluctuations in the vacuum have successfully explained many famous phenomena, such as the anomalous magnetic moment\cite{schwinger1948quantum}, Lamb shift\cite{lamb1947fine,bethe1947electromagnetic}, and Casimir forces\cite{casimir1948attraction,bordag2001new,milton2004casimir,plunien1986casimir}. In particular, quantum-fluctuation-related effects could be enhanced or modified by confining light modes in a small cavity\cite{PhysRevA.43.398,PhysRevA.103.033709,PhysRevA.92.063830,PhysRevA.91.063814}. Indeed, physicists have come to appreciate using quantum fluctuations in cavities to tailor the properties of materials or molecules \cite{hubener2021engineering,schlawin2022cavity}. For example, quantum fluctuations in cavities can mediate interactions between electrons, leading to the enhancement of superconductivity\cite{sentef2018cavity,curtis2019cavity,schlawin2019cavity,PhysRevResearch.4.013101,PhysRevLett.125.217402}, the breakdown of the quantization of Hall conductance\cite{ciuti2021cavity,appugliese2022breakdown,rokaj2022polaritonic}, or new phases of matter\cite{HDeng2002,PhysRevLett.109.053601,Byrnes2014,PhysRevLett.127.167201,PhysRevLett.123.207402,PhysRevLett.125.143603,rokaj2022free}. Furthermore, pioneering works have shown that quantum fluctuations in cavities can modify activation barriers and even affect chemical reaction rates significantly\cite{PhysRevLett.116.238301,flick2017atoms,galego2017many,galego2019cavity,schafer2019modification,altman2021quantum,PhysRevB.103.165412}.

However, quantum fluctuations in the vacuum (or a trivial cavity) preserve parity symmetry (PS) and are unable to induce an access of chirality in chemical reactions. Incorporating PS breaking into the quantum fluctuations is essential for selecting chirality. Several studies have demonstrated the influence of symmetry breaking on phenomena induced by quantum fluctuations, such as chirality-dependent Casimir forces\cite{salam1994maxwell,jenkins1994retarded,butcher2012casimir,barcellona2017enhanced, PhysRevB.91.235115,PhysRevA.97.032507,jiang2019chiral,farias2020casimir,hoye2020casimir,safari2020medium},
dissipationless Casimir viscosity\cite{jiang2019axial}, and band gap generation\cite{kibis2011band,wang2019cavity,PhysRevB.104.L081408,jiang2023engineering}. To highlight the combined power of the symmetry breaking and quantum fluctuations, Wilczek and one of us proposed a general framework, showing that symmetry breaking can be transmitted from materials to their vicinity via vacuum quantum fluctuations. The vacuum in proximity to a symmetry-broken material is referred as its quantum atmosphere\cite{jiang2019quantum}. 

\begin{figure}[!htb]
\centering
\includegraphics[height=2.32cm, width=8.6cm, angle=0]{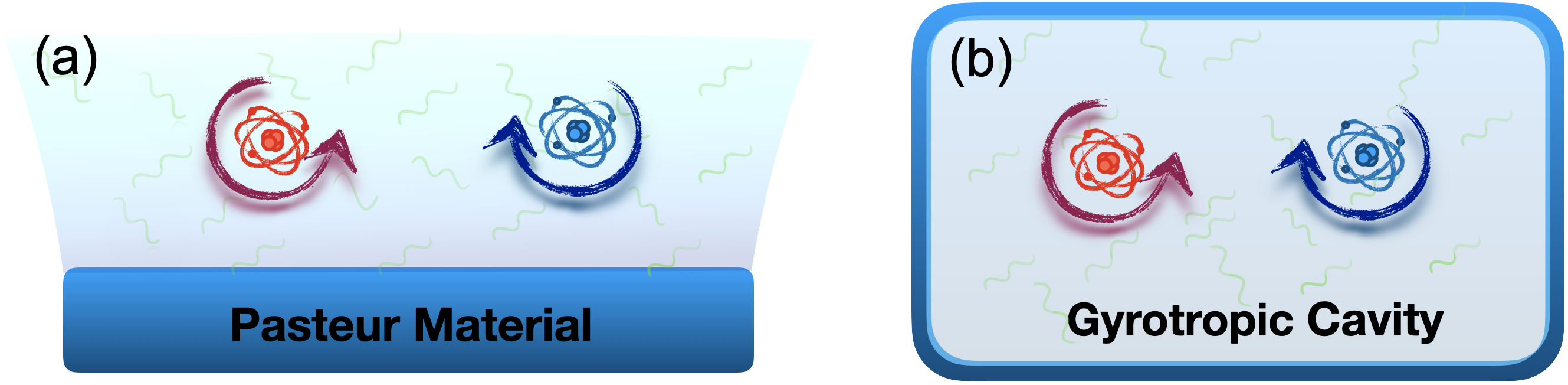}
\caption{A pair of enantiomers  (red and blue colors refer to opposite chirality)  immersed in PS-broken quantum fluctuations proximate to a Pasteur material (a) and in a gyrotropic cavity (b).}
\label{fig1}
\end{figure}

In this paper, we show that the quantum fluctuations proximate to a PS-broken material can induce a notable energy difference between a chiral molecule and its mirror-image enantiomer. The energy difference shifts the activation barrier in chemical reactions, resulting in an excess of one particular chirality over the other. 
We propose two setups for exploring this effect: one involving a Pasteur material and the other utilizing a gyrotropic cavity (Fig. \ref{fig1}). 
Notice that gyrotropic cavities, which permit the existence of only one circular polarization of photonic mode, have been realized experimentally using patterned chiral mirrors \cite{fedotov2006mirror,voronin2022single,baranov2023toward}, spin-orbital coupled materials \cite{gautier2022planar}, and topological photonic materials \cite{semnani2020spin,owens2022chiral}.

The paper is organized as follows: We first review the theoretical model of chiral molecules based on the Born-Oppenheimer (BO) approximation\cite{BOapproxima}.
Second, we identify a chiral energy shift that can characterize chiral ensembles in a PS-broken vacuum. Through first-principle calculation, we uncover a sizable energy disparity of  a pair of enantiomers. Finally, we evaluate the chemical reaction rate and obtain a notable chirality production rate, amply illustrating the selective power of symmetry-broken quantum fluctuations.

{\it BO approximation and the model of chiral molecules.---}
The Hamiltonian of a molecule contains three parts: the energy of nucleus, the energy of electrons, and the interaction energy between the electrons and the nucleus:
\begin{equation}
	\widehat H_{mol}=\widehat T_{n}+\widehat V_{n} (\hat{\bold R})+\widehat T_{ el}+\widehat V_{el}(\hat{\bold r})+\widehat V_{ n-el}(\hat{\bold R},\hat{\bold r}).
\end{equation}
Here, $\widehat T$ and $\widehat V$ denote the kinetic energy and Coulomb interactions, respectively; subindices $n$ and $el$ represent nucleus and electrons, respectively; variables $\left\{\bold R_1, \bold R_2,\dots, \bold R_{N_1}\right\}$ and $\left\{\bold r_1, \bold r_2,\dots, \bold r_{N_2}\right\}$ stand for the positions of $N_1$ nucleus and $N_2$ electrons, respectively. 

BO approximation is based on the fact that the kinetic energy of electrons is much larger than the nucleus. The energy-scale separation allows one to treat the electron movement first while the nuclear positions are regarded as fixed parameters. According to this scenario, the BO approximation is a two-step procedure\cite{BOapproxima,henriksen2018theories}:
\begin{enumerate}[(1)]
	\item
	Solve the electronic Hamiltonian
	$
	\widehat H_{el}
	=\widehat T_{el}+\widehat V_{el}+\widehat V_{n-el}(\bold R,\hat{\bold r}),
	$
	whose $i$-th eigenenergy and eigenstate are functions of $\bold R$, denoted as $E_{ el, i}(\bold R)$ and $|\phi_{i}(\bold R)\rangle$ $(i=0,1,2...)$, respectively.
	\item 
	Promoting $\mathbf{R}$ to be an operator yields an $\mathbf{\hat R}$-dependent electronic Hamiltonian $\widehat H_{el}(\mathbf{\hat R})=\sum_i E_{el,i}(\mathbf{\hat R})|\phi_i(\mathbf{\hat R})\rangle\langle \phi_i (\mathbf{\hat R})|$.  \end{enumerate}
The potential energy surface (PES) for the $i$th electronic energy level is defined as 
\begin{equation}
	{{\widehat V}_i}( {{\mathbf{\hat R}}}) \equiv  {{\widehat V}_{n}}( {{\mathbf{\hat R}}} ) +E_{el,i}(\mathbf{\hat R})|\phi_i(\mathbf{\hat R})\rangle\langle \phi_i (\mathbf{\hat R})|.
\end{equation} 
Because of the well-separated PESs, the transition between different PESs can be ignored, and the effective Hamiltonian for the $i$th PES reduces to ${{\widehat H}^{(i)}_{mol}} = {{\widehat T}_n}({{\mathbf{\hat R}}})+ {{\widehat V}_{i}}({{\mathbf{\hat R}}})$. 
We will focus on the shift of the ground-state PES of a chiral molecule induced by PS-broken quantum fluctuations. 

To characterize chiral molecules, we introduce a parity operator $\mathcal{J}_P$ and a rotation operator $\mathcal{J}_R$. For a chiral molecule, any rotation of nuclear configurations could not bring the electronic Hamiltonian identical to its enantiomer, i.e., ${{\widehat H}_{el}}(\mathcal{J}_R\mathbf{\hat R})\neq {{\widehat H}_{el}} (\mathcal{J}_P \mathbf{\hat R})$ for all $ \mathcal J_R \in {\rm SO(3)}$. By contrast, for achiral molecules, there always exists a rotation operation that can bring the electronic Hamiltonian identical to its parity counterpart. 

{\it Energy shift of chiral molecules in vacuum.---}
We consider a molecule interacting with a nearby symmetry-broken material through vacuum quantum EM fluctuations. Within the BO approximation, and the total Hamiltonian consists the electronic Hamiltonian $\widehat H_{el}$ of the molecule, the vacuum EM Hamiltonian $\widehat H_{em}$, the electron-EM interaction $\widehat H_{el-em}$, and the material-EM interaction $\widehat H_{mat-em}$:
\begin{eqnarray}
	\widehat H=\widehat H_{el}+\widehat H_{el-em}+\widehat H_{em}+\widehat H_{mat-em}.
\end{eqnarray} 
Tracing out the material degrees of freedom $|\psi_{mat}\rangle$ yields an effective Hamiltonian $H_{sb}$ that embodies the symmetry-breaking information of the material:
\begin{eqnarray}
	\widehat H_{sb}=\widehat H_{em}+{\rm Tr}_{mat}\left(\widehat H_{mat-em}\right)
\end{eqnarray}
where ${\rm Tr}_{mat}$ denotes the trace of the matrix with respect to the states of the material.
Since $\widehat H_{sb}$ encodes all the material's influence on the vacuum EM fields, a simplified Hamiltonian can be employed to address the interaction between the molecules and the symmetry-broken material:
\begin{eqnarray}
	\widehat H=\widehat H_{el}+\widehat H_{el-em}+\widehat H_{sb}.
\end{eqnarray}
When considering the interactions between molecules and EM fields, the important Fourier components of EM fields are those whose frequencies are on the order of atomic frequencies or less. Since the corresponding wavelength is much larger than the size of the molecule, one could employ the interaction Hamiltonian in the multipolar coupling scheme and long-wavelength approximation, describing the interaction of a molecule with the EM fields:
$\widehat H_{el-em}=-\mathbf{\widehat d}\cdot \mathbf{\widehat E}-\mathbf{\widehat m}\cdot \mathbf{\widehat B}$, where $\bold d$ and $\bold m$ denote the electric and magnetic dipoles of the molecule, respectively. This multipolar coupling Hamiltonian has been extensively studied in the literature\cite{power1983,salam1994maxwell,salam1997pra,jenkins1994retarded,butcher2012casimir}.
Electromagnetic quantum fluctuations renormalize the electronic energy levels, yielding a shift of the ground-state energy (we use the unit $\hbar=c=1$)
%\begin{widetext}
\begin{equation}
	\Delta E_0=-\sum_{i, I, F} p(I)
	\frac{\left|\left\langle\phi_i, F\left|\widehat{\bold{d}} \cdot \widehat{\bold{E}}+\widehat{\bold{m}} \cdot \widehat{\bold{B}}\right| \phi_0, I\right\rangle\right|^2}{E_{i 0}+\Omega_{F I}}.\nonumber
\end{equation}
%\end{widetext}
Here, $|\phi_i, I\rangle\equiv |\phi_i(\bold{R})\rangle \otimes |I\rangle$ is the direct product state of the $i$th PES's state $|\phi_i(\bold{R})\rangle$ and the photonic state $|I\rangle$. $p (I)=e^{-\beta E_I}/Z_{ph}$ is the thermal probability of the initial photonic state $|I\rangle$. $E_{i0}$ is the energy gap between the $i$th PES and the lowest PES.
$\Omega_{FI}=\Omega_F-\Omega_I$ is the energy gap between the photonic states $|F\rangle$ and $|I\rangle$.

The ground-state energy shift includes two parts: The first contribution
\begin{eqnarray}\label{achiralenterm}
	\Delta E_{0}^{a\chi} =
	- \sum\limits_{i,I,F} p (I) \frac{\left| \mathbf d_{0i}\cdot \mathbf E_{IF}\right|^2+\left| \mathbf m_{0i}\cdot \mathbf B_{IF}\right|^2}{E_{i0} + \Omega_{FI}}
\end{eqnarray}
is achiral, because it remains invariant under a parity operation (i.e., $\bold m\rightarrow \bold m$ and $\bold d\rightarrow -\bold d$).
By contrast, the second contribution 
\begin{equation}\label{chiralenshift}
	\Delta E_{0}^{\chi}=-\sum_{i,I,F}p(I)\frac{2\,{\rm Re }\left[\left(\bold d_{0i}\cdot \bold E_{IF}\right)\left(\bold m_{i0}\cdot \bold B_{FI}\right)\right]}{E_{i0}+\Omega_{FI}}
\end{equation}
reverses its sign under the parity operation and is called {\bf chiral energy shift}. The transition matrices are defined as
$\bold d_{0i}=\langle \phi_0|\widehat{\bold d}|\phi_i\rangle$, $\bold m_{i0}=\langle \phi_i|\widehat{\bold m}|\phi_0\rangle$, $\bold E_{IF}=\langle I | \bold{\widehat E}| F\rangle$, and $\bold B_{FI}=\langle F |\bold{\widehat B}| I\rangle$. 
%Similar formulas with spatially decayed features have been used molecular response theory

The chiral energy shift arises for a molecule (regardless of being chiral or not) with finite electronic and magnetic dipoles. However, to calculate the average chiral energy shift of an isotropic ensemble (liquid or gas), one should integrate over all orientations. And that makes the key difference: The average chiral energy shift vanishes for achiral ensembles but remains finite for chiral ensembles (see the proof \footnotemark[1]):
\begin{equation}\label{aces}
	\langle \Delta E_{0}^{\chi}\rangle
	= -\sum_{i,I,F}\frac{2\,p(I)\,{\rm Re}\left[\mathcal R_{i0}\left(\bold E_{IF}\cdot \bold B_{FI}\right)\right]}{3\left(E_{i0}+\Omega_{FI}\right)},
\end{equation}
where $\mathcal R_{i0}\equiv\left(\bold d_{0i}\cdot \bold m_{i0}\right)$, the imaginary part of which is called rotatory strength\cite{craig1998molecular,barron2009molecular}. In what follows, we will evaluate the chiral energy shift of chiral molecules induced by two types of PS-broken quantum fluctuations. 

{\it Chiral molecules above a Pasteur material.---}
The EM response of a Pasteur material is governed by the constitutive relations 
$\bold D=\epsilon \bold E-i\kappa \bold H$ and $\bold B=\mu\bold H+i\kappa\bold E$,
where, following Landau's convention, $\bold D$ and $\bold B$ ($\bold E$ and $\bold H$) are called electric and magnetic induction (field); $\epsilon=\epsilon_r \epsilon_0$ ($\mu=\mu_r\mu_0$) are the permittivity (permeability) of the Pasteur material\cite{lindell1994electromagnetic}; the parameter $\kappa$ characterizes the strength of PS breaking. While the Casimir-Polder forces between a chiral molecule and a Pasteur material have been nicely investigated\cite{butcher2012casimir}, we focus on the spectral change. 

\begin{figure}[!htb]
	\centering
	\includegraphics[height=3.cm, width=8.8cm, angle=0]{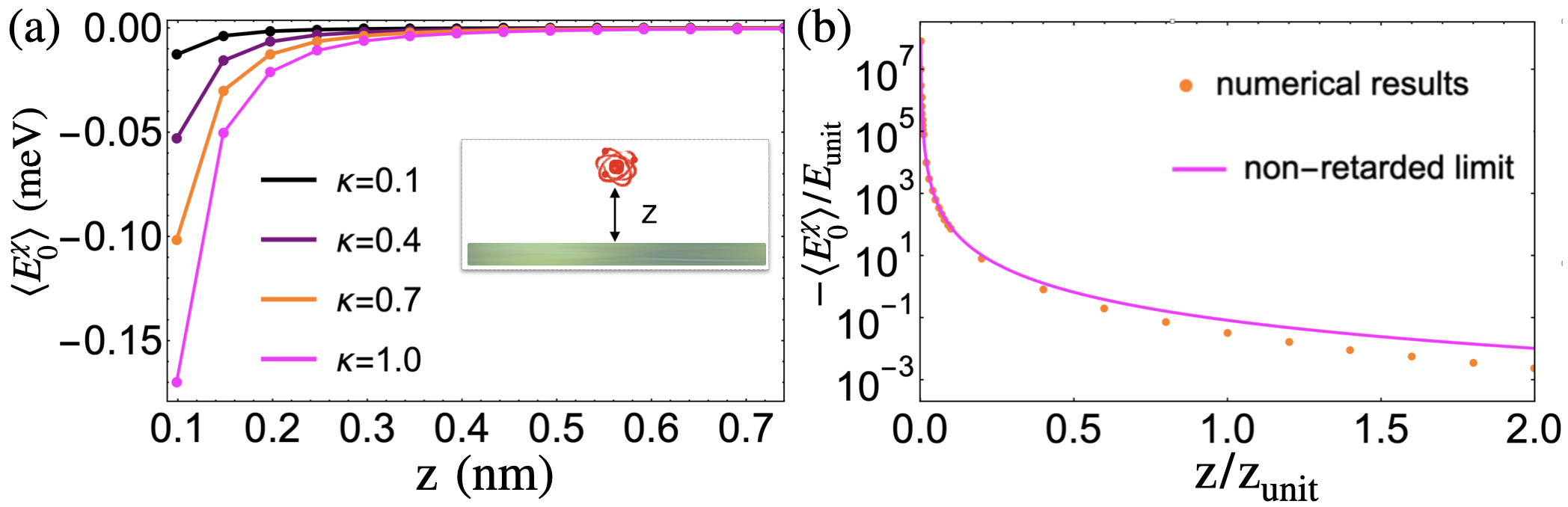}
	\caption{(a) Numerically calculation of average chiral energy shift for different parameter $\kappa$. We set $E_{10} = 2 \, {\rm eV}$ and $\operatorname{Im}\mathcal R_{10}=0.1 e a_0 \mu_B$ with $a_0$ and $\mu_B$ the Bohr radius and Bohr magnetic moment, respectively. (b) The analytical nonretarded result works for $z\ll z_{\rm unit}=1/E_{10}$; the energy unit is defined as $E_{\rm unit}={\mu_0 \operatorname{Im}\mathcal R_{10} E_{10}^3}/{3\pi^2}$; $\kappa = 0.4$.  We set ${\varepsilon _r} = {\mu _r} = 1$.}
	\label{fig2}
\end{figure}

We can alternatively express Eq. \eqref{aces} in terms of Green's functions\footnotemark[1]\footnotetext{See the supplemental materials for details including references [\onlinecite{abrikosov2012methods,buhmann2013dispersion,galego2015cavity,kowalewski2016non,tully2000perspective,henkelman2000a,henkelman2000b}].}:  
	\begin{equation}\label{aces_gf}
		\langle\Delta E_0^{\chi}\rangle =
		\frac{{2{\mu _0}}}{{3\pi }}\sum\limits_i {\text{Im}}{\mathcal{R}_{i0}}{\int_0^{ + \infty } {\frac{{{\xi ^2}d\xi }}{{E_{i0}^2 + {\xi ^2}}}{\text{Tr}}\nabla  \times {\mathbf{G}}\left( {i\xi } \right)} } 
	\end{equation}
with $\bold G$ the Green's function of EM fields, which includes the free-space contribution $\bold G_{free}$ and the symmetry-breaking contribution from Pasteur material $\bold G_{sb}$. Similar formulas Eqs.\eqref{achiralenterm}, \eqref{chiralenshift}, and \eqref{aces_gf} have been used to derive chiral-surface-molecule interactions and chiral-surface-mediated interactions previously\cite{butcher2012casimir,barcellona2017enhanced}.
%Change: mention a previous result consistent with ours
While the free-space contribution embodies no symmetry-breaking information and leads to identical energy shift for enantiomers, the symmetry-breaking contribution differentiates chiral enantiomers. One can obtain $\bold G_{sb}$ from the reflection coefficients of the Pasteur plate, namely $r_{ss}$, $r_{sp}$, $r_{ps}$, and $r_{pp}$, where the subindices indicate the polarization of the incoming and reflected EM wave\footnotemark[1]. We numerically calculate the distance-dependent energy shift, and compare it with the analytical nonretarded limit $\frac{{{\mu _0}\left( {{r_{sp}} - {r_{ps}}} \right)}}{{48i\pi {z^3}}}\operatorname{Im} {\mathcal{R}_{eg}}$ (Fig. \ref{fig2}). Here, $z$ is the distance between the chiral molecule and Pasteur plate, and ${\rm Im} \mathcal R_{10}$ is the rotatory strength considering the molecular ground state and its first excited state. 

{\it Chiral molecules in gyrotropic cavities.---}
We study chiral molecules in a PS-broken gyrotropic cavity. 
The Hamiltonian for the cavity photonic modes reads
$\hat H_{qa}=\sum_n\Omega_n \left(\hat a_n^\dagger \hat a_n+\frac{1}{2}\right)$,
where $\hat a_n^\dagger$ is the creation operator of a photonic mode $n$ and frequency $\Omega_n$. In terms of creation and annihilation operators, the vector potential is
\begin{equation}
	\bold{\hat A}(\bold r)=\sum_n g_n \left[\bold  A_n(\bold r)\hat a_n+\bold  A_n^*(\bold r) \hat a_n^\dagger\right],
\end{equation}
where the coupling strength $g_n=\sqrt{{1}/{2\epsilon_0 \Omega_n V_{{\rm eff}}}}$; and
$\bold  A_n(\bold r)=e^{i\bold k_n\cdot \bold r}\mathbf{\hat e}$ represents the EM wave of polarization $\mathbf{\hat e}$. We assume a general polarization $\mathbf{\hat e}=\bold{\hat e}_R+i \bold{\hat e}_I$ with real $\mathbf{\hat e}_R$ and $\mathbf{\hat e}_I$, which encodes the symmetry-broken information. 
Substituting the EM fields operators (Coulomb gauge) 
\begin{equation}
	\begin{aligned}
		\hat{\bold E}(\bold r)&=i \sum_n g_n \Omega_n \left[\bold A_n(\bold r)\hat a-\bold A_n^*(\bold r) \hat a^\dagger\right], ~\text{and}\\
		\hat{\bold B}(\bold r)&=i \sum_n g_n \left[\boldsymbol\nabla\times\bold  A_n(\bold r)\hat a+\boldsymbol\nabla \times\bold  A_n^*(\bold r) \hat a^\dagger\right]
	\end{aligned}
\end{equation}
into Eq. \eqref{aces} yields the chiral energy shift\footnotemark[1]
\begin{equation}\label{cesgc}
	\langle\Delta E_0^{\chi}\rangle=\sum_n\frac{4g_n^2\Omega_n^2}{3}\bold{\hat e_k}\cdot\left(\bold{\hat e}_R\times\bold{\hat e}_I\right) \sum_i \frac{{\rm Im} \mathcal R_{i0}}{E_{i0}+ \Omega_n}.
\end{equation}
Here, the factor $\bold{\hat e_k}\cdot\left(\bold{\hat e}_R\times\bold{\hat e}_I\right)$ flips sign under parity operation and implies the chirality of the cavity mode. $\bold{\hat e_k}\cdot\left(\bold{\hat e}_R\times\bold{\hat e}_I\right)=\pm \frac{1}{2}$ for a right(left)-handed circularly polarized mode, whereas it vanishes in a linearly polarized one. 
Let us estimate the magnitude of the chiral energy shift with promising experimental parameters. 
Considering two-level systems (ground state $|g\rangle$ and excited state $|e\rangle$),
%we estimate the chiral energy shift in a positive chirality cavity:  
we estimate the chiral energy shift in a left-handed gyrotropic cavity
\begin{equation}\label{chiralEn}
	\langle\Delta E_0^{\chi}\rangle = -\frac{4\pi \alpha}{3}\sum_n \left(\frac{{\rm Im}\mathcal R_{eg}}{e a_0\mu_B}\right)\left(\frac{a_0^3}{V_{\rm eff}}\right)\frac{ \Omega_n}{{\Delta E +  \Omega_n }}E_{\rm Ryd}
\end{equation}
where $\alpha$, $a_0$, and $\mu_B$ are the fine structure constant, Bohr radius, and Bohr magnetic moment, respectively. $\Delta E $ denotes the electronic energy gap; $\Omega_n$ is the frequency of the $n$th photonic mode; $E_{\rm Ryd}$ is the Rydberg energy; ${\rm Im}\mathcal R_{eg}$ is the molecular rotatory strength, and the effective mode volume of a cavity is defined as the ratio between the total field energy in the cavity divided by the field energy density at the molecular position, i.e., ${V_{\rm eff}} = \int_V {{d^3}r\varepsilon \left( {\mathbf{r}} \right){{\left| {{\mathbf{E}}\left( {\mathbf{r}} \right)} \right|}^2}}/{\epsilon_0\left| {{\mathbf{E}}\left( {{{\mathbf{r}}_m}} \right)} \right|^2}$. 
Setting ${\rm Im} \mathcal R_{eg} \approx 0.1\, e a_0\mu_B$, $\Delta E\approx 2\, {\rm eV}$, $\Omega_n=0.1 n \,{\rm eV}$ (ten modes $n=1,2...10$), and the smallest effective volume reachable in experiments $V_{\rm eff}\approx 0.2\, \text{nm}^3$, we obtain an experimental detectable chiral energy shift $\overline {\delta E_{0}^{\rm \chi}}  \approx  -0.06 \,{\rm meV}$. 

\begin{figure}[!htb]
	\centering
	\includegraphics[height=6.cm, width=8.6cm, angle=0]{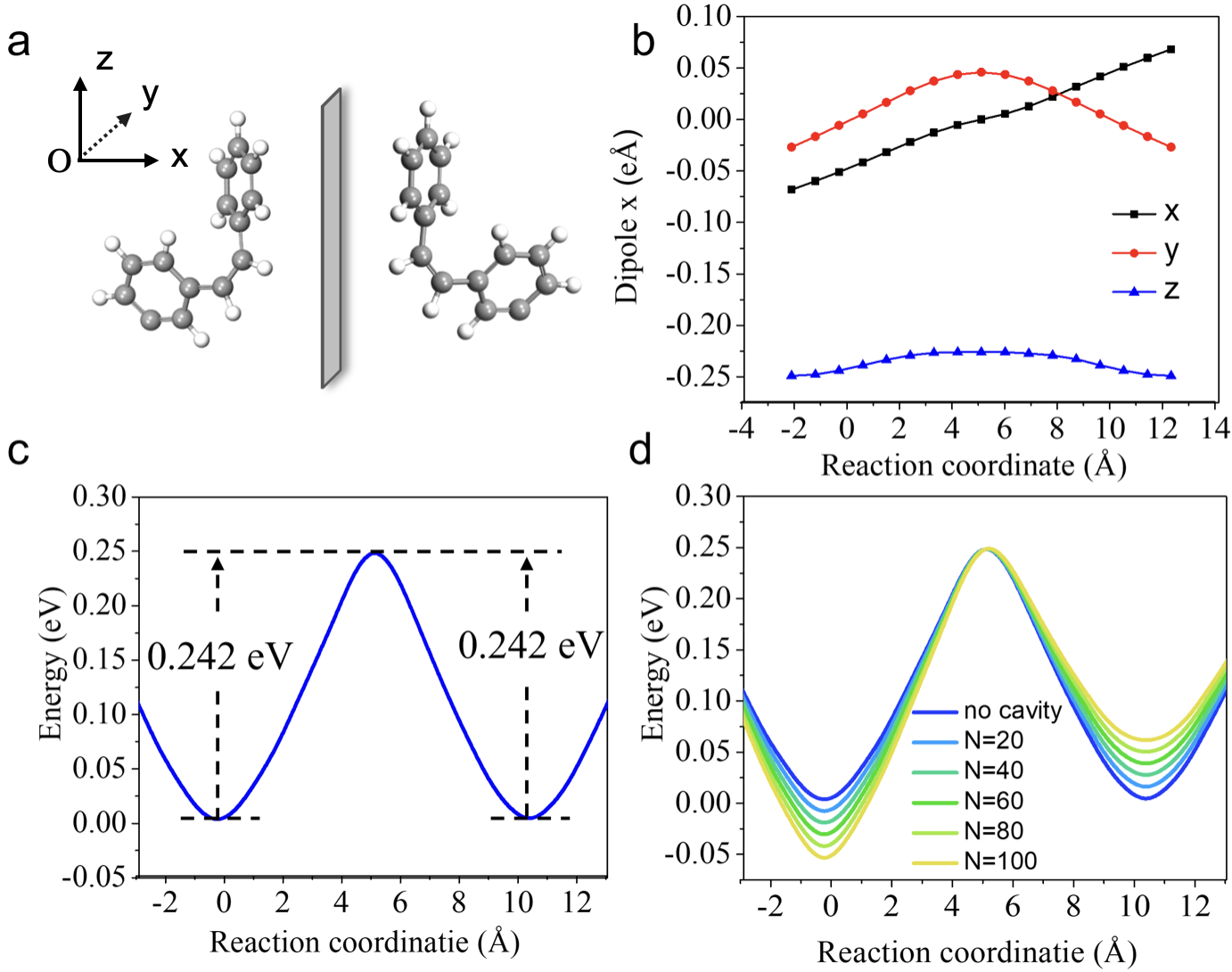}
	\caption{(a) Atomic structure of a chiral molecule and its enantiomer. Gray and white balls represent carbon and hydrogen atoms, respectively. (b) DFT calculated electric dipole as a function of the reaction coordinate. 
			The reaction coordinate is defined as a certain average of nuclei displacement, which captures the change of molecular configuration.
			The reaction coordinate is defined as a certain average displacement of nucleus, capturing the change of molecular configuration. (c) Bare molecular PES. The barrier is symmetric for left-handed and right-handed molecules. (d) Molecular PES versus the reaction coordinate induced by chiral energy shift [see Eq. \eqref{eq_Debye energy 1}]. Different colors correspond to different numbers of molecules.}
	\label{fig3}
\end{figure}

{\it  Collective enhancement.---}
Here, we explore how the chiral energy shift can be enhanced by collective effects. For this analysis, it is important to identify two types of terms in Eq. \eqref{chiralenshift}, 
commonly called the Debye term ($i=0$) and the London term ($i \neq 0$). 
While both terms contribute to the measurable spectral shift, only the Debye term can benefit from collective effect in a polarized ensemble of molecules (see below). Similar to Eq. \eqref{chiralenshift}, we calculate the Debye part of the chiral energy shift of all $N$ molecules at zero temperature\footnotemark[1]:
		\begin{eqnarray}	
			&&{\left. {\Delta E_{0,N}^\chi } \right|_{{\text{Debye}}}} \nonumber\hfill \\
			&&=  - \sum\limits_n {\frac{{2\operatorname{Re} \left[ {\left( {\sum\limits_{j = 1}^N {{\mathbf{d}}_{00}^{\left( j \right)} \cdot {{\mathbf{E}}_{n,01}^{(j)}}} } \right)\left( {\sum\limits_{j = 1}^N {{\mathbf{m}}_{00}^{\left( j \right)} \cdot {{\mathbf{B}}_{n,10}^{(j)}}} } \right)} \right]}}{{{\Omega _n}}}}  \nonumber \\
			&&=  - {N^2}\sum\limits_n {\frac{{2\operatorname{Re} \left[ {\left( {{\mathbf{d}}_{00}^{} \cdot {{\mathbf{E}}_{n,01}}} \right)\left( {{\mathbf{m}}_{00}^{} \cdot {{\mathbf{B}}_{n,10}}} \right)} \right]}}{{{\Omega _n}}}}  
		\end{eqnarray}
where $n$ labels the cavity modes, and $(j)$ represents the physical quantities of the $j$th molecule. The second equality follows that the molecules are polarized, ${\mathbf{d}}_{00}^{\left( j \right)} \cdot {{\mathbf{E}}_{n,01}^{(j)}} = {\mathbf{d}}_{00}^{} \cdot {{\mathbf{E}}_{n,01}},\,{\mathbf{m}}_{00}^{\left( j \right)} \cdot {{\mathbf{B}}_{n,10}^{(j)}} = {\mathbf{m}}_{00}^{} \cdot {{\mathbf{B}}_{n,10}}$. (A similar derivation was also given in \cite{galego2019cavity}.) 
	Modeling chiral cavity modes the same as before,
	the Debye term per molecule is
	\begin{equation}
		\begin{aligned}
			{\left. {\Delta E_0^\chi } \right|_{{\text{Debye}}}} =&  - N\sum\limits_n {\frac{{2\operatorname{Re} \left[ {\left( {{{\mathbf{d}}_{00}} \cdot {{\mathbf{E}}_{n,01}}} \right)\left( {{{\mathbf{m}}_{00}} \cdot {{\mathbf{B}}_{n,10}}} \right)} \right]}}{{{\Omega _n}}}}  \hfill \\
			=&  - N\sum\limits_n {g_n^2\Omega _n^{}\left( {{d_{00,x}}{m_{00,y}} - {d_{00,y}}{m_{00,x}}} \right)}  \hfill \\
		\end{aligned}
	\end{equation}
	
	To illustrate the opposite energy shift for a pair of enantiomers, we consider a concrete example --- an ensemble of chiral molecules named hydrogen-missing helicene [Fig. \ref{fig3}(a)]. Using density functional theory (DFT), we calculate the ground-state electric dipole moment of this molecule as it undergoes a transition from left-handed to right-handed configuration\footnotemark[1] [Fig. \ref{fig3}(b)]. Enantiomers are mirror images of each other across the $y$-$z$ plane, resulting in dipole moments of $\left( { \pm {d_{00,x}},{d_{00,y}},{d_{00,z}}} \right)$. The ground-state magnetic moment arises from an unpaired electron, which can be polarized in the $y$ direction. Consequently, the pair of enantiomers experience opposite chiral energy shifts 
	%Change: explain that London term is ignored
	(note that the London term does not scale with $N$ and can be ignored here \footnotemark[1]):
		\begin{equation}\label{eq_Debye energy 1}
			\Delta E_0^\chi\approx{\left. {\Delta E_0^\chi } \right|_{{\text{Debye}}}} =   \mp N\sum\limits_n {g_n^2\Omega _n^{}{d_{00,x}}{m_{00,y}}} 
	\end{equation}
Using the same parameters as given below Eq. \eqref{chiralEn} except with ${{\mathbf{d}}_{00}} = 0.2e{a_0}{{\mathbf{e}}_x}$ and ${{\mathbf{m}}_{00}} = {\mu _B}{{\mathbf{e}}_y}$, we find the magnitude $ \mp N \times 0.92 \, {\text{meV}}$. Our DFT calculations in Fig. \ref{fig3}(d) show that the chiral energy shift in a cluster of 100 molecules can be significantly enhanced. We have not considered the effect of intermolecular interactions\cite{jenkins1994retarded,salam1994maxwell,safari2020medium}, which may influence the collective enhancement quantitatively by either strengthening or weakening the molecular alignment.

\begin{figure}[!htb]
	\centering
	\includegraphics[height=3.66cm, width=7.2cm, angle=0]{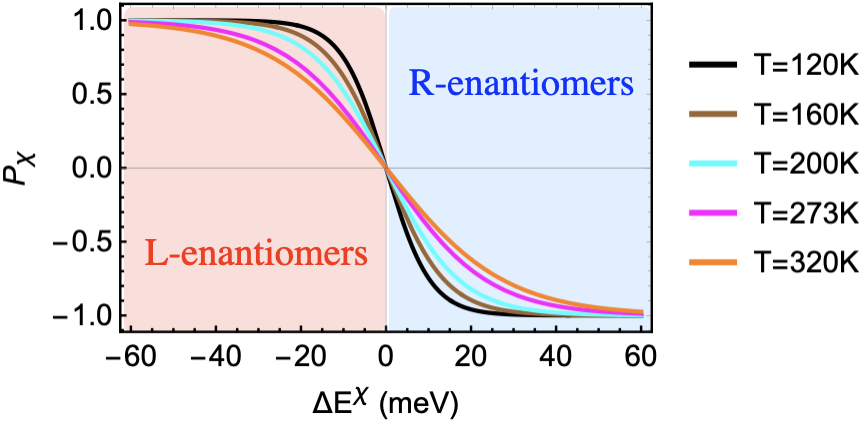}
	\caption{Chirality-selective rate as a function of the chiral energy shift at different temperatures. Upon reversing the chiral energy shift, opposite chirality is selected. 
	}
	\label{fig4}
\end{figure}

Finally, we evaluate the impact of chiral energy shift in chemical reactions. 
According to the collision theory, chemical reaction rates depends on the activation energy $E_a$---a minimum amount of energy that the reactants need to overcome to form products. The chemical reaction rates can be approximately calculated by the Arrhenius equation $k=A \exp(-\beta E_{a})$ ($k$ is the rate constant and  $E_{a}$ is the activation energy)\cite{laidler1984development}. 
We define a {\it chirality-selective rate} to characterize the reaction-rate difference between chiral enantiomers\footnotemark[1]: 
\begin{equation}\label{eq_rate}
		P_\chi=\frac{k_L-k_R}{k_L+k_R}=\frac{1-\exp{\left(-2\beta{ \Delta E_0^{\chi} }\right)}}{1+\exp{\left(-2\beta{ \Delta E_0^{\chi} }\right)}}
\end{equation} 
with $k_L$ ($k_R$) denoting the chemical reaction rates for the \textit{L}(\textit{D}) enantiomer. In Supplemental Material \footnotemark[1], we use transition state theory to justify the above result. One could tune the sign of the chiral energy shift to select desired chirality in chemical reactions (see Fig. \ref{fig4}). 

{\it Summary.---}
We studied the effect of quantum fluctuations on the spectra of chiral molecules. Our research demonstrates that PS-broken quantum fluctuations can induce a chirality-dependent shift in the ground-state energy. We predicted a significant rate of chirality selection for chiral enantiomers in a gyrotropic cavity. Since chirality selection is related to the rotational strength or the ground-state dipole moments (generally nonzero), our findings have broad applicability and are not limited to any specific molecular model. We remark that our logic, content, and proposals differ fundamentally from the recent papers discussing chirality discrimination\cite{riso2023strong,vu2023enhanced}. 

The authors gratefully acknowledge helpful discussions and suggestions from anonymous referees. We are sponsored by Pujiang Talent Program 21PJ1405400, TDLI starting up grant, Jiaoda 2030 program WH510363001, and Innovation Program for Quantum Science and Technology Grant No.2021ZD0301900.%

\bibliography{refchiral.bib}

\newpage

\clearpage
\begin{widetext}
\appendix

\begin{center}
\textbf{Supplemental Materials}
\end{center}

The supplemental materials are structured as follows:
\begin{itemize}
\item 
Section A discusses the average of chiral energy shift across all spatial orientations, as given by Equation (8).
\item 
Sections B and C are dedicated to the scattering Green's function method for calculating the chiral energy shift. The specific calculation for the Pasteur plate case can be found in Section C.
\item 
Section D presents the derivation of the chiral energy shift in a gyrotropic cavity, as given by Equation (12).
\item
Section E explores the effects of finite temperature on our results.
\item
Section F provides an explanation of the approximations utilized in the Arrhenius equation we employ.
\item
Section G contains information regarding the density functional theory calculation.
\end{itemize}

\section{Isotropic average of chiral energy shift}\label{sec_Iso Ave}

In this section, we delve deeper into the derivation of averaging the chiral energy shift over various molecular orientations. We provide the derivation of Equation (8) presented in the main text and explain the reason behind its vanishing value for achiral molecules.
Considering a molecule with nuclear positions denoted as $\mathbf{R}$, the chiral energy shift can be expressed as:
	\begin{equation}
		{\Delta E_0^\chi } =  - \sum\limits_{i,I,F} {{p_I}\frac{{2\operatorname{Re} \left( {{{\mathbf{d}}_{0i}}\left( {\mathbf{R}} \right) \cdot {{\mathbf{E}}_{IF}}} \right)\left( {{{\mathbf{m}}_{i0}}\left( {\mathbf{R}} \right) \cdot {{\mathbf{B}}_{FI}}} \right)}}{{{E_{i0}} + {\omega _{FI}}}}} 
	\end{equation}
 It is important to note that the matrix elements of electric and magnetic dipoles are dependent on the orientation of $\mathbf{R}$. In the case of isotropic samples of molecules, as discussed in the main text, it is necessary to take an average over all orientations. Therefore, the expression can be written as follows:
	\begin{equation}\label{eq_average ccp}
		\left\langle {\Delta E_0^\chi } \right\rangle   =  - \int_{{\mathcal{J}_R} \in {\text{SO}}\left( 3 \right)} {d{\mathcal{J}_R}\sum\limits_{i,I,F} {{p_I}\frac{{2\operatorname{Re} \left( {{{\mathbf{d}}_{0i}}\left( {{\mathcal{J}_R}{\mathbf{R}}} \right) \cdot {{\mathbf{E}}_{IF}}} \right)\left( {{{\mathbf{m}}_{i0}}\left( {{\mathcal{J}_R}{\mathbf{R}}} \right) \cdot {{\mathbf{B}}_{FI}}} \right)}}{{{E_{i0}} + {\omega _{FI}}}}} } 
	\end{equation}
	But what are ${{\mathbf{d}}_{0i}}\left( {{\mathcal{J}_R}{\mathbf{R}}} \right)$ and ${{\mathbf{m}}_{0i}}\left( {{\mathcal{J}_R}{\mathbf{R}}} \right)$? Define a unitary representation of SO(3), $\hat{J}_R$, acting on electronic position eigenstates as ${\hat{J}_R}\left| {\mathbf{r}} \right\rangle  = \left| {{\mathcal{J}_R}{\mathbf{r}}} \right\rangle $. Then, using the fact that electronic Hamiltonian (see the discussion of BO approximation in the main text) is unchanged under rotating both the nuclear and electronic positions, $\hat J_R^{ - 1}H\left( {{\mathcal{J}_R}{\mathbf{R}}} \right){{\hat J}_R} = H\left( {\mathbf{R}} \right)$, one can verify that $\left| {{\phi _i}\left( {{\mathcal{J}_R}{\mathbf{R}}} \right)} \right\rangle  = {{\hat J}_R}\left| {{\phi _i}\left( {\mathbf{R}} \right)} \right\rangle $. Moreover, for matrix elements of electric dipole moment,
	\begin{equation}
	\begin{aligned}
		{{\mathbf{d}}_{ij}}\left( {\mathbf{R}} \right) =& \sum\nolimits_{\alpha  \in \{ {\text{nuclei}}\} } {{Z_\alpha }e{{\mathbf{R}}_\alpha }}  - \left\langle {{\phi _i}\left( {\mathbf{R}} \right)} \right|\sum\nolimits_{\beta  \in \{ {\text{electrons}}\} } {e{{\hat {\mathbf{r}}}_\beta }} \left| {{\phi _j}\left( {\mathbf{R}} \right)} \right\rangle  \hfill \\
		{{\mathbf{d}}_{ij}}\left( {{\mathcal{J}_R}{\mathbf{R}}} \right) =& \sum\nolimits_{\alpha  \in \{ {\text{nuclei}}\} } {{Z_\alpha }e{\mathcal{J}_R}{{\mathbf{R}}_\alpha }}  - \left\langle {{\phi _i}\left( {\mathbf{R}} \right)} \right|\hat J_R^{ - 1}\sum\nolimits_{\beta  \in \{ {\text{electrons}}\} } {e{{\hat {\mathbf{r}}}_\beta }} {{\hat J}_R}\left| {{\phi _j}\left( {\mathbf{R}} \right)} \right\rangle  \hfill \\
		=& {\mathcal{J}_R}{{\mathbf{d}}_{ij}}\left( {\mathbf{R}} \right) \hfill 
	\end{aligned} 
	\end{equation}
	Similarly, magnetic moment $\displaystyle {{\mathbf{m}}_{ij}}\left( {\mathbf{R}} \right) = \left\langle {{\phi _i}\left( {\mathbf{R}} \right)} \right|\sum\nolimits_{\beta  \in \{ {\text{electrons}}\} } {\frac{{ - e}}{{2{m_e}}}{{{\mathbf{\hat r}}}_\beta } \times {{{\mathbf{\hat p}}}_\beta }} \left| {{\phi _j}\left( {\mathbf{R}} \right)} \right\rangle $ satisfies 
	\begin{equation}
		{{\mathbf{m}}_{ij}}\left( {{\mathcal{J}_R}{\mathbf{R}}} \right) = {\mathcal{J}_R}{{\mathbf{m}}_{ij}}\left( {\mathbf{R}} \right)
	\end{equation}
	Also, for parity operator introduced in the main text, ${\mathcal{J}_P}$, which is a spatial inversion around the center of a molecule, one can prove that
	\begin{equation}
		{{\mathbf{d}}_{ij}}\left( {{\mathcal{J}_P}{\mathbf{R}}} \right) =  - {{\mathbf{d}}_{ij}}\left( {\mathbf{R}} \right),{{\mathbf{m}}_{ij}}\left( {{\mathcal{J}_P}{\mathbf{R}}} \right) = {{\mathbf{m}}_{ij}}\left( {\mathbf{R}} \right)
	\end{equation}

	Using these definitions, \eqref{eq_average ccp} becomes
	\begin{equation}
		\begin{aligned}
			\left\langle {\Delta E_0^\chi } \right\rangle   =&  - \int_{{\mathcal{J}_R} \in {\text{SO}}\left( 3 \right)} {d{\mathcal{J}_R}\sum\limits_{i,I,F} {{p_I}\frac{{2\operatorname{Re} \left[ {\left( {{\mathcal{J}_R}{{\mathbf{d}}_{0i}} \cdot {{\mathbf{E}}_{IF}}} \right)\left( {{\mathcal{J}_R}{{\mathbf{m}}_{i0}} \cdot {{\mathbf{B}}_{FI}}} \right)} \right]}}{{{E_{i0}} + {\omega _{FI}}}}} }  \hfill \\
			=&  - \frac{2}{3}\sum\limits_{i,I,F} {{p_I}\frac{{\operatorname{Re} \left[ {\left( {{{\mathbf{d}}_{0i}} \cdot {{\mathbf{m}}_{i0}}} \right)\left( {{{\mathbf{E}}_{IF}} \cdot {{\mathbf{B}}_{FI}}} \right)} \right]}}{{{E_{i0}} + {\omega _{FI}}}}}  \hfill 
		\end{aligned} 
	\end{equation}
	This is eq.8 in the main text. Note that ${{{\mathbf{d}}_{0i}} \cdot {{\mathbf{m}}_{i0}}}$ changes sign under parity operation, $\left\langle {\Delta E_0^\chi \left( {{\mathcal{J}_P}{\mathbf{R}}} \right)} \right\rangle  =  - \left\langle {\Delta E_0^\chi \left( {\mathbf{R}} \right)} \right\rangle $. However, as mentioned in the main text, for an achiral molecule, there is a specific rotation $\mathcal{J}_{{R_0}}$, s.t. $H\left( {{\mathcal{J}_{{R_0}}}{\mathbf{R}}} \right) = H\left( {{\mathcal{J}_P}{\mathbf{R}}} \right)$ and $\left\langle {\Delta E_0^\chi \left( {{\mathcal{J}_{{R_0}}}{\mathbf{R}}} \right)} \right\rangle  = \left\langle {\Delta E_0^\chi \left( {{\mathcal{J}_P}{\mathbf{R}}} \right)} \right\rangle $. So, 
	\[
	\begin{aligned}
		\left\langle {\Delta E_0^\chi \left( {{\mathcal{J}_P}{\mathbf{R}}} \right)} \right\rangle   =& \int_{{J_R} \in {\text{SO}}\left( 3 \right)} {d{\mathcal{J}_R}\delta E_0^{{\text{ch}}}\left( {{\mathcal{J}_R}{\mathcal{J}_P}{\mathbf{R}}} \right)}  \hfill \\
		=& \int_{{J_R} \in {\text{SO}}\left( 3 \right)} {d{\mathcal{J}_R}\delta E_0^{{\text{ch}}}\left( {{\mathcal{J}_{{R_0}}}{\mathcal{J}_R}{\mathbf{R}}} \right)}  \hfill \\
		=& \left\langle {\Delta E_0^\chi \left( {\mathbf{R}} \right)} \right\rangle   \hfill 
	\end{aligned} 
	\]
	In the second equality we use that parity and rotation commute. Therefore, average chiral energy shift indeed vanishes for achiral molecules.
	
	\section{Chiral Casimir-Polder energy in electromagnetic Green's function}\label{app_b}
	
	In this section, we derive the explicit expression for chiral Casimir-Polder energy using electromagnetic Green's function, i.e., Eq.(9) in the main text, on which the next section is based.
	
	According to Eq.(7) of the main text, the chiral energy shift is,
	\begin{equation}
		\Delta E_0^\chi \left( {{{\mathbf{r}}_M}} \right)  =  - \sum\limits_{i,I,F} {p\left( I \right)\frac{{2\operatorname{Re} \left[ {{{\mathbf{d}}_{0i}} \cdot {{\mathbf{E}}_{IF}}\left( {{{\mathbf{r}}_M}} \right) \otimes {{\mathbf{B}}_{IF}}\left( {{{\mathbf{r}}_M}} \right) \cdot {{\mathbf{m}}_{i0}}} \right]}}{{{E_{i0}} + {E_{FI}}}}} 
	\end{equation}
	where we combine vectors $\mathbf{E}$ and $\mathbf{B}$ to be a tensor. 
	
	Electromagnetic Green's function is a useful tool for studying this problem. Roughly, it is both the Green's function for the equation of motion of vector potential in $\phi  = 0$ gauge, obeying $\left( { - \frac{{{\omega ^2}}}{{{c^2}}}{\varepsilon _r}\left( {{\mathbf{r}},\omega } \right) + \nabla  \times \frac{1}{{{\mu _r}\left( {{\mathbf{r}},\omega } \right)}}\nabla  \times } \right){\mathbf{G}}\left( {{\mathbf{r}},{\mathbf{r}}',\omega } \right) = \delta \left( {{\mathbf{r}} - {\mathbf{r}}'} \right)$ (where Pasteur parameter $\kappa$ is zero for simplicity), and the retarded response function of vector potential (times $ - \mu _0^{ - 1}$) (For an introduction to electromagnetic Green's function, see Ref.[S1,S2]). To use Green's function in the following derivation, we need the analytic structure of correlation function. The frequency domain retarded response function of operators ${\widehat O_{1,2}}$ is ${G^R_{12}}\left( \omega  \right) = {Z^{ - 1}}\sum\limits_{I,F} {{O_{1,IF}}{O_{2,FI}}\frac{{{e^{ - \beta {E_I}}} - {e^{ - \beta {E_F}}}}}{{\omega  + {E_{IF}} + i0 + }}} $. Generalized imaginary part can be defined as $\operatorname{Im} '{G^R}(\omega) = \frac{1}{2i} \left[{ G_{12}^R(\omega) - G_{21}^{R\; *}(\omega) }\right]$, and thus
	\begin{equation}
		\operatorname{Im} '{G^R}\left( \omega  \right) =  - \pi \sum\limits_{I,F} {p\left( I \right){O_{1,IF}}{O_{2,FI}}\left( {1 - {e^{ - \beta \omega }}} \right)\delta \left( {\omega  - {E_{FI}}} \right)} 
	\end{equation}
	Therefore, we have the following expression of the generalized imaginary part of the correlation function of electric and magnetic fields
	\begin{equation}
		 - \frac{1}{\pi }\int_{ - \infty }^{ + \infty } {d\omega \frac{{\operatorname{Im} '{\mathbf{G}}_{em}^R\left( {{\mathbf{r}},{\mathbf{r}}',\omega } \right)}}{{\left( {1 - {e^{ - \beta \omega }}} \right)\left( {{E_{i0}} + \omega } \right)}}}  = \sum\limits_{I,F} {p\left( I \right)\frac{{{{\mathbf{E}}_{IF}}\left( {\mathbf{r}} \right) \otimes {{\mathbf{B}}_{IF}}\left( {{\mathbf{r}}'} \right)}}{{{E_{i0}} + {E_{FI}}}}} 
	\end{equation}
	The $em$ correlation function can be related to the Green function (of vector potential) through $\operatorname{Im} '{\mathbf{G}}_{em}^R\left( {{\mathbf{r}},{\mathbf{r}}',\omega } \right) = i{\mu _0}\omega \operatorname{Im} '{\mathbf{G}}\left( {{\mathbf{r}},{\mathbf{r}}',\omega } \right) \times \overleftarrow \nabla  '$[S4]. So we can write the chiral energy shift in a compact form,
	\begin{equation}\label{eq_gf ccp1}
		\Delta E_0^\chi \left( {{{\mathbf{r}}_M}} \right) = - \sum\limits_i {\int_{ - \infty }^{ + \infty } {\frac{{{\mu _0}\omega d\omega }}{{\pi \left( {1 - {e^{ - \beta \omega }}} \right)\left( {{E_{i0}} + \omega } \right)}}{\text{2Im Tr}}\left( {{{\mathbf{m}}_{i0}}{{\mathbf{d}}_{0i}} \cdot \operatorname{Im} '{\mathbf{G}}\left( {{{\mathbf{r}}_M},{{\mathbf{r}}_M},\omega } \right) \times \overleftarrow \nabla  '} \right)} }  
	\end{equation}
	
	We study an easier case here. We assume zero temperature, i.e. only consider vacuum field contribution, and isotropic molecular sample. The latter is to say that we replace the tensor like ${\mathbf{dm}}$ with its average over all directions $\left\langle {{\mathbf{dm}}} \right\rangle  = \frac{{{\mathbf{d}} \cdot {\mathbf{m}}}}{3}{\mathbf{I}}$ (see section \ref{sec_Iso Ave}). Note that $\operatorname{Im} '{\mathbf{G}}\left( {{\mathbf{r}},{\mathbf{r}}',\omega } \right) = \left[ {{\mathbf{G}}\left( {{\mathbf{r}},{\mathbf{r}}',\omega } \right) - {{\mathbf{G}}^T}\left( {{\mathbf{r}}',{\mathbf{r}},\omega } \right)} \right]/2i$, so $\operatorname{Im} '{\mathbf{G}}\left( {{{\mathbf{r}}_M},{{\mathbf{r}}_M},\omega } \right) \times \overleftarrow \nabla  ' = \frac{1}{{2i}}\left\{ {{\mathbf{G}}\left( {{{\mathbf{r}}_M},{{\mathbf{r}}_M},\omega } \right) \times \overleftarrow \nabla  ' + {{\left[ {\nabla  \times {{\mathbf{G}}^*}\left( {{{\mathbf{r}}_M},{{\mathbf{r}}_M},\omega } \right)} \right]}^T}} \right\}$, and \eqref{eq_gf ccp1} can be written as
	\begin{equation}\label{eq_gf ccp2}
		\left\langle {\Delta E_0^\chi \left( {{{\mathbf{r}}_M}} \right)} \right\rangle  = \sum\limits_i {\int_0^{ + \infty } {\frac{{{\mu _0}\omega d\omega }}{{3\pi \left( {{E_{i0}} + \omega } \right)}}{\text{Re}}\left[ {{\mathcal{R} _{i0}}\left( {{\text{Tr}}\left( {{\mathbf{G}} \times \overleftarrow \nabla  '} \right) + {\text{Tr}}\left( {\nabla  \times {{\mathbf{G}}^*}} \right)} \right)} \right]} } 
	\end{equation}
	where $\mathcal R_{i0}\equiv\left(\bold d_{0i}\cdot \bold m_{i0}\right)$
	
	{  To treat the ${\nabla  \times {{\mathbf{G}}^*}\left( {{{\mathbf{r}}_M},{{\mathbf{r}}_M},\omega } \right)}$ term, we use the Schwarz reflection principle, 
	\begin{equation}\label{eq_SchRef}
		{{\mathbf{G}}^*}\left( {{\mathbf{r}},{{\mathbf{r}}' },\omega } \right) = {\mathbf{G}}\left( {{\mathbf{r}},{{\mathbf{r}}' }, - {\omega ^*}} \right)
	\end{equation}
	which is actually a general property of response functions. We also need Onsager reciprocal relation for electromagnetic Green's function, which leads to ${\text{Tr}}\left( {\nabla  \times {\mathbf{G}}\left( {{{\mathbf{r}}_M},{{\mathbf{r}}_M}} \right)} \right) =  - {\text{Tr}}\left( {{\mathbf{G}}\left( {{{\mathbf{r}}_M},{{\mathbf{r}}_M}} \right) \times \overleftarrow \nabla  '} \right)$. Combining them, \eqref{eq_gf ccp2} is now
	\begin{equation}
		\begin{aligned}
			\left\langle {\Delta E_0^\chi \left( {{{\mathbf{r}}_M}} \right)} \right\rangle  &= \sum\limits_i {\int_0^{ + \infty } {\frac{{{\mu _0}\omega d\omega }}{{3\pi \left( {{E_{i0}} + \omega } \right)}}{\text{Re}}\left[ {{\mathcal{R} _{i0}}\left( { - {\text{Tr}}\left[ {\nabla  \times {\mathbf{G}}\left( \omega  \right)} \right] + {\text{Tr}}\left[ {\nabla  \times {{\mathbf{G}}}\left( { - \omega } \right)} \right]} \right)} \right]} }  \hfill \\
			&=  - \sum\limits_i {\frac{{i{\mu _0}\operatorname{Im} {\mathcal{R} _{i0}}}}{{3\pi }}\left( {\int_0^{ + \infty } {\frac{{\omega d\omega }}{{{E_{i0}} + \omega }}{\text{Tr}}\left[ {\nabla  \times {\mathbf{G}}\left( \omega  \right)} \right]}  - \int_0^{ - \infty } {\frac{{\omega d\omega }}{{{E_{i0}} - \omega }}{\text{Tr}}\left[ {\nabla  \times {{\mathbf{G}}}\left( \omega  \right)} \right]} } \right)}  \hfill \\ 
		\end{aligned} 
	\end{equation}
	The second equality can be checked by simply expanding all $\operatorname{Re}$ and $\operatorname{Im}$ part and using Eq.\eqref{eq_SchRef}.
	
	By using the contour integration techniques, the integral interval can be transformed to the upper imaginary axis, $\omega  = i\xi $ with $\xi$ from $0$ to $+ \infty$. The expression becomes
	\begin{equation}\label{eq_gf ccp3}
		\begin{aligned}
			\left\langle {\Delta E_0^\chi \left( {{{\mathbf{r}}_M}} \right)} \right\rangle   &= \sum\limits_i {\frac{{i{\mu _0}\operatorname{Im} {\mathcal{R} _{i0}}}}{{3\pi }}\int_0^{ + \infty } {\xi d\xi \left( {\frac{{{\text{Tr}}\left[ {\nabla  \times {\mathbf{G}}\left( {i\xi } \right)} \right]}}{{{E_{i0}} + i\xi }} - \frac{{{\text{Tr}}\left[ {\nabla  \times {{\mathbf{G}}}\left( {i\xi } \right)} \right]}}{{{E_{i0}} - i\xi }}} \right)} }  \hfill \\
			&=  \sum\limits_i {\frac{{2{\mu _0}\operatorname{Im} {\mathcal{R}_{i0}}}}{{3\pi }}\int_0^{ + \infty } {\frac{{{\xi ^2}d\xi }}{{E_{i0}^2 + {\xi ^2}}}{\text{Tr}}\left[ {\nabla  \times {\mathbf{G}}\left( {i\xi } \right)} \right]} }  \hfill \\ 
		\end{aligned}  
	\end{equation}
	which proves Eq.(9) in the main text. 
	
	Note that a well-defined ${\kappa\left( {\omega } \right)}$ also follows a "Schwarz reflection principle" ${\kappa ^*}\left( \omega  \right) =  - \kappa \left( { - {\omega ^*}} \right)$ such that the Green's function satisfies \eqref{eq_SchRef}. To make things easier, in Appendix.\ref{app_c}, we simply model the Pasteur parameter to be $\operatorname{sgn} \left( {\operatorname{Re} \omega } \right)\kappa $, where $\kappa$ is real. Now, when doing continuation from the positive/negative real-axis to the upper imaginary-axis, the results are different; we denote the Green's function in the half plane $\operatorname{Re} \omega > 0$ or $<0$ as ${{\mathbf{G}}_{+\kappa}\left( {i\xi } \right)}$ or ${{\mathbf{G}}_{-\kappa}\left( {i\xi } \right)}$, i.e., ${\mathbf{G}}\left( {\operatorname{Re} \omega  > 0} \right) = {{\mathbf{G}}_{ + \kappa }}\left( \omega  \right)$, ${\mathbf{G}}\left( {\operatorname{Re} \omega  < 0} \right) = {{\mathbf{G}}_{ - \kappa }}\left( \omega  \right)$. They are related by ${{\mathbf{G}}_{ - \kappa }}\left( { - {\omega ^*}} \right) = {\mathbf{G}}_{ + \kappa }^*\left( \omega  \right)$, with which \eqref{eq_gf ccp3} should be revised:
	\begin{equation}\label{eq_gf ccp4}
		\begin{aligned}
			\left\langle {\Delta E_0^\chi \left( {{{\mathbf{r}}_M}} \right)} \right\rangle  =& \sum\limits_i {\frac{{i{\mu _0}\operatorname{Im} {\mathcal{R}_{i0}}}}{{3\pi }}\int_0^{ + \infty } {\xi d\xi \left( {\frac{{{\text{Tr}}\left[ {\nabla  \times {{\mathbf{G}}_{ + \kappa }}\left( {i\xi } \right)} \right]}}{{{E_{i0}} + i\xi }} - \frac{{{\text{Tr}}\left[ {\nabla  \times {{\mathbf{G}}_{ - \kappa }}\left( {i\xi } \right)} \right]}}{{{E_{i0}} - i\xi }}} \right)} }  \hfill \\
			=&  - \sum\limits_i {\frac{{2{\mu _0}\operatorname{Im} {\mathcal{R}_{i0}}}}{{3\pi }}\int_0^{ + \infty } {\frac{{\xi d\xi }}{{E_{i0}^2 + {\xi ^2}}}\left( {{E_{i0}}\operatorname{Im} {\text{ Tr}}\left[ {\nabla  \times {{\mathbf{G}}_{ + \kappa }}\left( {i\xi } \right)} \right] - \xi \operatorname{Re} {\text{ Tr}}\left[ {\nabla  \times {{\mathbf{G}}_{ + \kappa }}\left( {i\xi } \right)} \right]} \right)} }  \hfill \\ 
		\end{aligned} 
	\end{equation}
    }
	
	\section{Chiral Casimir-Polder effect near a half-space Pasteur material}\label{app_c}
	
	Based on Appendix.\ref{app_b}, here we discuss the effect of a chiral molecule near a half-space Pasteur material, as shown in Fig.2 in the main text, with numerical results and a non-retarded limit asymptotic forms. In this section, we will calculate Green's function on the positive real-axis and then do continuation to the upper imaginary-axis, and instead of ${{{\mathbf{G}}_{ + \kappa }}\left( {\omega } \right)}$ we write ${{\mathbf{G}}\left( {\omega } \right)}$ for simplicity (see the last paragraph of Appendix.\ref{app_b}).
	
	Note that we only consider quantum atmosphere contribution to the Green's function here [S5], which is the scattering Green's function. The expression is
	\begin{equation}\label{eq_GF1}
		\begin{gathered}
			{\mathbf{G}}\left( {{\mathbf{r}},{\mathbf{r}}',\omega } \right) = \int {{d^2}{k_\parallel }\frac{i}{{8{\pi ^2}{k_ \bot }}}\left( {{{\mathbf{a}}_{k + ,s}}\left( {\mathbf{r}} \right){\mathbf{a}}_{k - ,s}^*\left( {{\mathbf{r}}'} \right){r_{ss}} + {{\mathbf{a}}_{k + ,p}}\left( {\mathbf{r}} \right){\mathbf{a}}_{k - ,p}^*\left( {{\mathbf{r}}'} \right){r_{pp}}} \right.}  \hfill \\
			\left. { + {{\mathbf{a}}_{k + ,s}}\left( {\mathbf{r}} \right){\mathbf{a}}_{k - ,p}^*\left( {{\mathbf{r}}'} \right){r_{sp}} + {{\mathbf{a}}_{k + ,p}}\left( {\mathbf{r}} \right){\mathbf{a}}_{k - ,s}^*\left( {{\mathbf{r}}'} \right){r_{ps}}} \right) \hfill 
		\end{gathered} 
	\end{equation}
	$k_\parallel$ is the wave vector projection in $xOy$ plane and ${k_ \bot } = \sqrt {{{\left( {\omega /c} \right)}^2} - k_\parallel ^2} $ with $\operatorname{Im} k_\bot \le 0$. ${{\mathbf{a}}_{k \pm ,\sigma }}\left( {\mathbf{r}} \right) = {{\mathbf{e}}_{\sigma  \pm }}{e^{i\left( {{{\mathbf{k}}_\parallel } \cdot {\mathbf{r}} \pm {k_ \bot }z} \right)}}$ with $s$ and $p$-wave polarization unit vectors ${{\mathbf{e}}_{s \pm }} = {{\mathbf{e}}_{{{\mathbf{k}}_\parallel }}} \times {{\mathbf{e}}_z}$ and ${{\mathbf{e}}_{p \pm }} = \frac{c}{\omega }\left( {{k_\parallel }{{\mathbf{e}}_z} \mp {k_ \bot }{{\mathbf{e}}_{{{\mathbf{k}}_\parallel }}}} \right)$. 
	
	$r_{ss}$ and $r_{pp}$ are $s$ and $p$-wave reflection coefficients, and $r_{sp(ps)}$ is the coefficient of $p$($s$) to $s$($p$)-wave reflection process, which are[S3] opposite when Pasteur parameter $\kappa \neq 0$, $r_{sp}= -r_{ps}$.
	\begin{equation}
		{r_{sp\left( {ps} \right)}} =  \pm \frac{{2i{\eta _0}\eta c}}{\Delta }\left( {{c_ + } - {c_ - }} \right)
	\end{equation}
	Here, $\eta  = \sqrt {\mu /\varepsilon } $ (${\eta _0} = \sqrt {{\mu _0}/{\varepsilon _0}} $) is the impedance of material (vaccum); $c = {k_ \bot }/k$ is the $\cos$ of incident angle; ${c_ \pm } = \cos {\theta _ \pm } = \sqrt {k_ \pm ^2 - k_\parallel ^2} /{k_ \pm }$ are two transmission angles, with ${k_ \pm } = \sqrt {{\varepsilon _r}{\mu _r}} k\left( {1  \pm {\kappa _r}} \right)$; ${\kappa _r} = \kappa /\sqrt {{\varepsilon _r}{\mu _r}} $ is the relative Pasteur parameters, which is normally between $\left[ { - 1,1} \right]$; and $\Delta  = \left( {\eta _0^2 + {\eta ^2}} \right)c\left( {{c_ + } + {c_ - }} \right) + 2{\eta _0}\eta \left( {{c^2} + {c_ + }{c_ - }} \right) $. If $\kappa= 0$, then $r_{ps}$ and $r_{sp}$ vanish; and if $\kappa$ change sign $r_{ps}$ and $r_{sp}$ also change signs. Define $r_{sp} = ir$, with $r = r(c)$ a function of incident angle cosine $c$.
	
	Take the curl of $\mathbf{G}$ is equal to acting ${\mathbf{k}} \times ...$ to it, such as
	\[
	\begin{aligned}
		\nabla  \times {{\mathbf{a}}_{k + ,s}}\left( {\mathbf{r}} \right){\mathbf{a}}_{k - ,s}^*\left( {{\mathbf{r}}'} \right){r_{ss}} &= i{{\mathbf{k}}_ + } \times {{\mathbf{a}}_{k + ,s}}\left( {\mathbf{r}} \right){\mathbf{a}}_{k - ,s}^*\left( {{\mathbf{r}}'} \right){r_{ss}} \hfill \\
		&=  - ik{{\mathbf{a}}_{k + ,p}}\left( {\mathbf{r}} \right){\mathbf{a}}_{k - ,s}^*\left( {{\mathbf{r}}'} \right){r_{ss}} \hfill 
	\end{aligned} 
	\]
	If we take the trace, since $s$ and $p$ polarization is orthogonal, it vanishes. Therefore, only the last two terms in \eqref{eq_GF1} contribute to ${\text{Tr}}\operatorname{Im} \left( {\nabla  \times {\mathbf{G}}} \right)$. So the cross reflections $r_{sp(ps)}$, or $\chi$ and $\kappa$, are necessary in the chiral CP effect. Finally we have
	\begin{equation}
		\begin{aligned}
			&{\text{Tr}} \left( {\nabla  \times {\mathbf{G}}\left( {{\mathbf{r}},{\mathbf{r}},\omega } \right)} \right) \hfill \\
			=& \int {{d^2}{k_\parallel }\frac{i}{{8{\pi ^2}{k_ \bot }}}{e^{2i{k_ \bot }z}}\left[ { - ik{{\mathbf{e}}_{k + ,p}} \cdot {{\mathbf{e}}_{k - ,p}}\left( r_{sp} \right) + ik{{\mathbf{e}}_{k + ,p}} \cdot {{\mathbf{e}}_{k - ,p}}\left( -r_{sp} \right)} \right]}  \hfill \\
			=& \int {{d^2}{k_\parallel }\frac{k}{{8{\pi ^2}{k_ \bot }}}{e^{2i{k_ \bot }z}}\left[ {\frac{{k_\parallel ^2 - k_ \bot ^2}}{{{k^2}}}  r_{sp} - \left( { - r_{sp} } \right)} \right]}  \hfill \\
			=& \frac{i}{{4{\pi ^2}{\omega ^2}}}\int {{d^2}{k_\parallel }\frac{{{e^{2i\sqrt {1 - {{\left( {{k_\parallel }/\omega } \right)}^2}} \omega z}}}}{{\sqrt {1 - {{\left( {\frac{{{k_\parallel }}}{\omega }} \right)}^2}} }}{r (c) }k_\parallel ^2}   \hfill 
		\end{aligned} 
	\end{equation}
	In imaginary frequency, this becomes
	\begin{equation}
		\begin{aligned}
			&{\text{Tr}} \left( {\nabla  \times {\mathbf{G}}\left( {{\mathbf{r}},{\mathbf{r}},i\xi } \right)} \right) \hfill \\
			=&   - \frac{{{i}}}{{4{\pi ^2}{\xi ^2}}}\int {{d^2}{k_\parallel }\frac{{{e^{ - 2\sqrt {1 + {{\left( {{k_\parallel }/\xi } \right)}^2}} \xi z/c}}}}{{\sqrt {1 + {{\left( {\frac{{{k_\parallel }}}{\xi }} \right)}^2}} }}{r(c')}k_\parallel ^2}   \hfill \\
			=&   - \frac{{{i}}}{{2\pi {\xi ^2}}}\int_0^{ + \infty } {{k_\parallel }d{k_\parallel }\frac{{{e^{ - 2\sqrt {1 + {{\left( {{k_\parallel }/\xi } \right)}^2}} \xi z/c}}}}{{\sqrt {1 + {{\left( {\frac{{{k_\parallel }}}{\xi }} \right)}^2}} }}{r(c')}k_\parallel ^2}   \hfill 
		\end{aligned} 
	\end{equation}
	Here $c$ in $r$ is replaced by $c'$, because of changing $\omega \rightarrow i\xi$:
	\begin{equation}
		c = \sqrt {1 - {{\left( {\frac{{{k_\parallel }}}{\omega }} \right)}^2}}  \to \sqrt {1 + {{\left( {\frac{{{k_\parallel }}}{\xi }} \right)}^2}}  = c'
	\end{equation}
	Change the integral variable to $c'={\sqrt {1 + {{\left( {{k_\parallel }/\xi } \right)}^2}} }$,
	\begin{equation}
		{\text{Tr}}\left( {\nabla  \times {\mathbf{G}}\left( {{\mathbf{r}},{\mathbf{r}},i\xi } \right)} \right) =  - \frac{{i{\xi ^2}}}{{2\pi }}\int_1^{ + \infty } {dc'{e^{ - 2c'\xi z}}r\left( {c'} \right)}  
	\end{equation}
	Substitute it into \eqref{eq_gf ccp3} and change the integral variable $\xi  \to \xi z$, note that now ${\text{Tr}}\nabla  \times {\mathbf{G}}$ only have imaginary part.
	\begin{equation}\label{eq_gf hs}
		\left\langle {\Delta E_0^\chi \left( {z} \right)} \right\rangle  = \frac{{{\mu _0}}}{{3{\pi ^2}{z^2}}}\sum\limits_i {{\omega _{i0}}\operatorname{Im} {\mathcal{R} _{i0}}\int_0^{ + \infty } {\frac{{{x^3}dx}}{{{{\left( {{\omega _{i0}}z} \right)}^2} + {x^2}}}\int_1^{ + \infty } {dc'{e^{ - 2xc'}}\left( {c{'^2} - 1} \right)r\left( {c'} \right)} } } 
	\end{equation}
	
	In the following we consider the case of $z \to 0+$, or the non-retarded limit. 
	
	\subsection{Non-retarded limit}
	
	In the non-retarded limit, i.e. $a={\omega _{i0}}z/c \ll 1$, the integral in \eqref{eq_gf hs} is
	\[
	\int_0^{ + \infty } {\frac{{{x^3}dx}}{{{a^2} + {x^2}}}\int_1^{ + \infty } {dc'{e^{ - 2xc'}}\left( {c{'^2} - 1} \right) r } } 
	\]
	$\frac{1}{{{a^2} + {x^2}}}$ is only significant near $0+$, so we replace ${x^3}\int_1^{ + \infty } {dc'{e^{ - 2xc'}}r''\left( {c{'^2} - 1} \right)} $ with its limit when approaching $0+$,
	\[
	\begin{aligned}
		&{\lim _{x \to 0 + }}{x^3}\int_0^{ + \infty } {dc'{e^{ - 2xc'}}r\left( {c'} \right)\left( {c{'^2} - 1} \right)}  \hfill \\
		=& r\left( { + \infty } \right){\lim _{x \to 0 + }}{x^3}\int_0^{ + \infty } {dc'{e^{ - 2xc'}}\left( {c{'^2} - 1} \right)}  \hfill \\
		=& r\left( { + \infty } \right){\lim _{x \to 0 + }}{e^{ - 2x}}\left( {1 + 2x} \right) \hfill \\
		=& r\left( { + \infty } \right) \hfill \\ 
	\end{aligned} 
	\]
	Therefore, the $x$-integral is
	\[
	\int_0^{ + \infty } {\frac{{{x^3}dx}}{{{a^2} + {x^2}}}\int_0^{ + \infty } {dc'{e^{ - 2xc'}}r (c') \left( {c{'^2} - 1} \right)} } \xrightarrow{{a \to 0 + }}\frac{{r\left( { + \infty } \right)}}{4}\int_0^{ + \infty } {\frac{{dx}}{{{a^2} + {x^2}}}}  = \frac{\pi }{{8a}}r\left( { + \infty } \right)
	\]
	Substitute it into Eq.(C7), we obtain (in SI unit. The $c$ here is the light velocity)
	\begin{equation}\label{eq_CCPnrLimit}
		\left\langle {\Delta E_0^\chi \left( {z} \right)} \right\rangle  = \frac{{{\mu _0}c}}{{24\pi {z^3}}}r\left( { + \infty } \right)\operatorname{Im} {\mathcal{R} _{eg}}
	\end{equation}
	where $r(+\infty)$ is valued in the limit $c' \to  + \infty $ or imaginary frequency $\xi  \to 0 + $. This is $\frac{{{\mu _0}c\left( {{r_{sp}} - {r_{ps}}} \right)}}{{48i\pi {z^3}}}\operatorname{Im} {\mathcal{R}_{eg}}$, which has been compared with the numerical result in Fig.2 of the main text.
	
	\section{Chiral Casimir-Polder effect in a gyrotropic cavity}
	
	This section focuses on the derivation from Eq.10 to Eq.12 in the main text. 
	
	At zero temperature, a gyrotropic cavity mode (labeled $n$) as described in the main text is at zero photon state. Since the only involved intermediate states in eq.8 of main text is one photon state, eq.8 gives chiral energy shift due to this mode:
	\begin{equation}
		  - \sum\limits_{i \ne 0} {\frac{{2{\mkern 1mu} {\mkern 1mu} {\text{Re}}\;{\text{Tr}}\left[ {{\mathcal{R}_{i0}}\left( {{{\mathbf{E}}_{01}} \cdot {{\mathbf{B}}_{10}}} \right)} \right]}}{{3\left( {{E_{i0}} + {\Omega _n}} \right)}}} 
	\end{equation}
	Using eq.11, and $\left\langle 0 \right|\hat a\left| 1 \right\rangle  = 1$ and $\left\langle 1 \right|\hat a\left| 0 \right\rangle  = 0$, we obtain ${{\mathbf{E}}_{01}} \cdot {{\mathbf{B}}_{10}} = \frac{{g_n^2\Omega _n^2}}{c}{{\mathbf{e}}_{\mathbf{k}}} \times {\mathbf{e}} \cdot {{\mathbf{e}}^*} = \frac{{2ig_n^2\Omega _n^2}}{c}{{\mathbf{e}}_{\mathbf{k}}} \cdot {{\mathbf{e}}_R} \times {{\mathbf{e}}_I}$. Finally, summing over cavity modes $n$ gives eq.12:
	\begin{equation}
		\left\langle {\Delta E_0^\chi } \right\rangle  = \sum\limits_n {\frac{{4g_n^2\Omega _n^2}}{3}} {{{\mathbf{\hat e}}}_{\mathbf{k}}} \cdot \left( {{{{\mathbf{\hat e}}}_R} \times {{{\mathbf{\hat e}}}_I}} \right)\sum\limits_i {\frac{{{\text{Im}}{\mathcal{R}_{i0}}}}{{{E_{i0}} + {\Omega _n}}}} 
	\end{equation}
	
	\section{Chiral Casimir-Polder energy modified by temperature}
	
	{  Thus far, our discussions have mainly focused the zero-temperature scenario, where the uncoupled state is a direct product state of the molecular eigenenergy state and the photon state of zero occupation number. In what follows, we will show that the condition is not too far from reality. Our discussion will be based on the gyrotropic cavity setting, both the London-type and the Debye-type terms.}
	
	{\it London-type chiral energy shift.---}
	The argument includes two steps. Firstly, we examine the energy shift caused by cavity modes with $\Omega_n \ll E_{eg}$ and establish that this component only undergoes a negligible correction at finite temperatures due to the smallness of $\Omega_n$. Secondly, we consider the case of $\Omega_n \gtrsim E_{eg}$ and demonstrate that this part also experiences a minor correction.
 
 Eq.(8) in the main text, combined with the linearity of ${{\mathbf{\hat E}}}$ and ${{\mathbf{\hat B}}}$ with respect to photon operators, indicates that the energy shift of the molecule's ground states resulting from different optical modes is independent. Let us now calculate the contribution of a single mode when it is in a state of thermal equilibrium. For the mode $n$,
	\begin{equation}\label{eq_finiteT1}
		\begin{aligned}
			{\left\langle {\Delta E_0^\chi } \right\rangle _n} =&  - \sum\limits_{i \ne 0} {\frac{{2p\left( 0 \right)}}{3}\frac{{{\text{Re}}\;{\text{Tr}}\left[ {{\mathcal{R}_{i0}}\left( {{{\mathbf{E}}_{01}} \cdot {{\mathbf{B}}_{10}}} \right)} \right]}}{{\left( {{E_{i0}} + \hbar {\Omega _n}} \right)}}}  \hfill \\
			&- \sum\limits_{i \ne 0,I \ne 0} {\frac{{2p\left( I \right)}}{3}\left( {\frac{{{\text{Re}}\;{\text{Tr}}\left[ {{\mathcal{R}_{i0}}\left( {{{\mathbf{E}}_{I,I + 1}} \cdot {{\mathbf{B}}_{I + 1,I}}} \right)} \right]}}{{\left( {{E_{i0}} + \hbar {\Omega _n}} \right)}} + \frac{{{\text{Re}}\;{\text{Tr}}\left[ {{\mathcal{R}_{i0}}\left( {{{\mathbf{E}}_{I,I - 1}} \cdot {{\mathbf{B}}_{I - 1,I}}} \right)} \right]}}{{\left( {{E_{i0}} - \hbar {\Omega _n}} \right)}}} \right)}  \hfill \\ 
		\end{aligned} 
	\end{equation}
	Make use of eq.11,
	\[
	\begin{gathered}
		{{\mathbf{E}}_{I,I + 1}} = i{g_n}{\Omega _n}\sqrt {I + 1} {e^{i{\mathbf{k}} \cdot {\mathbf{r}}}}{\mathbf{\hat e}},\;{{\mathbf{B}}_{I + 1,I}} =  - i\frac{{{g_n}{\Omega _n}}}{c}\sqrt {I + 1} {e^{ - i{\mathbf{k}} \cdot {\mathbf{r}}}}{{{\mathbf{\hat e}}}_{\mathbf{k}}} \times {{{\mathbf{\hat e}}}^*} \hfill \\
		{{\mathbf{E}}_{I,I - 1}} =  - i{g_n}{\Omega _n}\sqrt I {e^{ - i{\mathbf{k}} \cdot {\mathbf{r}}}}{{{\mathbf{\hat e}}}^*},{\text{ }}{{\mathbf{B}}_{I - 1,I}} = i\frac{{{g_n}{\Omega _n}}}{c}\sqrt I {e^{i{\mathbf{k}} \cdot {\mathbf{r}}}}{{{\mathbf{\hat e}}}_{\mathbf{k}}} \times {\mathbf{\hat e}} \hfill \\ 
	\end{gathered} 
	\]
	Assume the factor ${{\mathbf{e}}_{\mathbf{k}}} \cdot {{\mathbf{e}}_R} \times {{\mathbf{e}}_I}$ takes maximum $\frac{1}{2}$ for simplicity. \eqref{eq_finiteT1} becomes
	\begin{equation}
		\begin{aligned}
			{\left\langle {\Delta E_0^\chi } \right\rangle _n} &= \frac{{2g_n^2\Omega _n^2}}{{3c}}\sum\limits_{i \ne 0,I} {p\left( I \right)\left( {\left( {I + 1} \right)\frac{{{\text{Im}}{\mathcal{R}_{i0}}}}{{{E_{i0}} + \hbar {\Omega _n}}} - I\frac{{\operatorname{Im} {\mathcal{R}_{i0}}}}{{{E_{i0}} - \hbar {\Omega _n}}}} \right)}  \hfill \\
			&= \left\langle {\Delta E_0^\chi } \right\rangle _n^{\left(  +  \right)} + {n_B}\left( {\beta {\Omega _n}} \right)\left( {\left\langle {\Delta E_0^\chi } \right\rangle _n^{\left(  +  \right)} - \left\langle {\Delta E_0^\chi } \right\rangle _n^{\left(  -  \right)}} \right) \hfill \\ 
		\end{aligned} 
	\end{equation}
	where ${n_B}\left( {\beta \omega } \right) = 1/\left( {{e^{\beta \omega }} - 1} \right)$ is Bose distribution and $\left\langle {\Delta E_0^\chi } \right\rangle _n^{\left(  \pm  \right)} = \frac{{2g_n^2\Omega _n^2}}{{3c}}\frac{{{\text{Im}}{\mathcal{R}_{i0}}}}{{{E_{i0}} \pm \hbar {\Omega _n}}}$ correspond to ground state molecule energy shift due to a emission and a subsequent absorption (absorption and a subsequent emission) of one virtual photon.
	
	$\left\langle {\Delta E_0^\chi } \right\rangle _n^{\left(  +  \right)}$ is just the zero temperature energy shift and ${n_B}\left( {\beta {\Omega _n}} \right)\left( {\left\langle {\Delta E_0^\chi } \right\rangle _n^{\left(  +  \right)} - \left\langle {\Delta E_0^\chi } \right\rangle _n^{\left(  -  \right)}} \right)$ is the finite temperature correction. For $\Omega_n$ much smaller than molecule energy level, $\Omega_n \ll E_{eg}$, the correction is small: 
	\begin{equation}\label{eq_finiteT2}
	\frac{{{{\left\langle {\Delta E_0^\chi } \right\rangle }_n}\left( T \right)}}{{{{\left\langle {\Delta E_0^\chi } \right\rangle }_n}\left( {T = 0} \right)}} = 1 - {n_B}\left( {\beta {\Omega _n}} \right)\frac{{2\hbar {\Omega _n}}}{{{E_{eg}} - \hbar {\Omega _n}}}
	\end{equation}
	To estimate the correction ${n_B}\left( {\beta {\Omega _n}} \right)\frac{{2\hbar {\Omega _n}}}{{{E_{eg}} - \hbar {\Omega _n}}}$, we use the same parameters as in the main text, $\Delta E\approx 2\, {\rm eV}$, $\Omega_n=0.1 n \,{\rm eV}$ (ten modes $n=1,2...10$), and temperature $400\,\mathrm{K}\approx 0.034\,\mathrm{eV}$. All the modes' corrections are less than $0.6\%$, which is negligible. We then conclude that at finite temperature, the energy shift due to the modes whose frequencies are much smaller than $E_{eg}$ is slightly less than that of zero temperature.
	
	How about the modes whose $\Omega_n \gtrsim E_{eg}$? The $\left\langle {\Delta E_0^\chi } \right\rangle _n^{\left(  -  \right)}$ seems diverge when $\Omega_n \to E_{eg}$, leading to a divergent finite temperature correction. But this is because non-degenerate perturbation theory is used. If $\Omega_n \sim E_{eg}$, the excited states $\left| {I+1,g} \right\rangle $ with $(I+1)$ ($I \le 0$) photons and ground state molecule and $\left| {I,e} \right\rangle $ with $I$ photons and excited state molecule are degenerate, so the energy shift of $\left| {n + 1,g} \right\rangle $ diverges, which caused a divergent $\left\langle {\Delta E_0^\chi } \right\rangle _n^{\left(  -  \right)}$. In degenerate perturbation theory, one can prove that the low energy eigenstate in this subspace only obtain a finite energy shift and $\left\langle {\Delta E_0^\chi } \right\rangle _n^{\left(  -  \right)}$ is now finite. Though the expression $\left\langle {\Delta E_0^\chi } \right\rangle _n^{\left(  -  \right)}$ of modes with $\Omega_n \gtrsim E_{eg}$ is more complicated, we don't care much about them as long as they are finite, since their contribution will be multiplied by a prefactor ${n_B}\left(\beta E  \right) = 1/\left( {{e^{\beta E}} - 1} \right)$ with $E$ the energy scale of these excited states. $E$ is at least $\gtrsim E_{eg} \sim 2.3 \times 10^{4} \text{K}$, much higher than normal temperature. We conclude that the energy shift due to these mode can be safely obtained by ${\left\langle {\Delta E_0^\chi } \right\rangle _n}\left( T \right) = \left\langle {\Delta E_0^\chi } \right\rangle _n^{\left(  +  \right)} + {n_B}\left( {\beta {\Omega _n}} \right)\left( {\left\langle {\Delta E_0^\chi } \right\rangle _n^{\left(  +  \right)} - \left\langle {\Delta E_0^\chi } \right\rangle _n^{\left(  -  \right)}} \right) \approx \left\langle {\Delta E_0^\chi } \right\rangle _n^{\left(  +  \right)} = {\left\langle {\Delta E_0^\chi } \right\rangle _n}\left( {T = 0} \right)$.
	
	Combining the two conclusions above, we can argue that the finite temperature chiral energy shift should be only slightly smaller than in the zero temperature case.
	
	{ {\it Debye-type chiral energy shift.---}
	To obtain the Debye-type version of Eq.\eqref{eq_finiteT2}, we simply replace $E_{eg}$ with $0$. It is important to note that in this case, there is no resonance in this case and $\left\langle {\Delta E_0^\chi } \right\rangle _n^{\left(  -  \right)}$ does not go to infinite. The ratio is simply
	\begin{equation}\label{eq_finiteT3}
		\frac{{{{\left\langle {\Delta E_0^\chi } \right\rangle }_n}\left( T \right)}}{{{{\left\langle {\Delta E_0^\chi } \right\rangle }_n}\left( {T = 0} \right)}} = 1 + 2 {n_B}\left( {\beta {\Omega _n}} \right)
	\end{equation}
Therefore, the finite temperature chiral energy shift is expected to be higher than the zero temperature one. With the previous given parameters, $\Omega_n=0.1 n \,{\rm eV}$ (ten modes $n=1,2...10$), and temperature $400\,\mathrm{K}\approx 0.034\,\mathrm{eV}$, the relative correction does not exceed $2{n_B}\left( {0.1/0.034} \right) = 11\% $.}
 
  {  \section{Approximations behind the rate equation (Arrhenius equation)}}
	
In this section, we explain the two approximations behind translating the Casimir-Polder energy shift to reaction rate difference, Eq.(17) and Fig.4. First, we use the ground state potential energy surface (PES) to describe the process of chemical reactions. Within the adiabatic approximation, the nucleus in a molecule move independently on distinct PESs. In this work, we assume this approximation and thus only the lowest/ground state PES is relevant. Secondly, we use the Arrhenius equation to estimate the rate difference between the enantiomers. Consequently, the rates are determined solely by the disparity in barrier heights. We analyze more on these two approximations below.
	
	The adiabatic approximation, or the lowest PES approximation, is valid as long as the non-adiabatic couplings between PESs are negligible. In real world, typically, an exception only happens when two PESs come close to each other, such as the conical intersection. In cavity chemistry, i.e., quantum chemistry involving "photonic degrees of freedom", this phenomenon is even more usual. When the cavity mode frequency equals the electronic energy level spacing, some PESs (or sometimes called polaritonic surfaces in this context) which would have been occasionally degenerate at certain points without light-matter coupling, will hybridize to give a new set of polaritonic surfaces when light-matter interaction is turned on. Between these new surfaces, the non-adiabatic couplings can be strong and the adiabatic approximation can break down [S6,S7]. This does not happen in what we focus on because we are studying the lowest surface labeled by zero photon and electronic ground state. One can estimate the order of magnitude of the non-adiabatic couplings to assess its importance. Derived similarly to the original work by Born and Oppenheimer, without the presence of things like conical intersections, the couplings between polaritonic surfaces labeled by photon numbers and electronic quantum numbers are as small as those between the bare molecule PESs and thus can generally be neglected. Roughly, one can estimate the non-adiabatic couplings as below. Our analysis is basically in the same spirit as that of Born and Oppenheimer's [S8], so the readers familiar with their results may want to skip the rest of this paragraph. We assume one electromagnetic field mode with frequency $\Omega$ and electric dipole interaction here for simplicity. The non-adiabatic coupling between two PESs labeled by $\alpha$ and $\beta$ is $\frac{1}{{2M}}\left( {2{{\mathbf{F}}_{ij}} \cdot {\mathbf{P}} + {G_{ij}}} \right)$, where $M$ is a typical nuclear mass, $\mathbf{P}$ is the nuclear momentum, and ${{\mathbf{F}}_{\alpha \beta }} = \left\langle {{\psi _\alpha }} \right|{\nabla _{\mathbf{R}}}\left| {{\psi _\beta }} \right\rangle $ and ${G_{\alpha \beta }} = \left\langle {{\psi _\alpha }} \right|\nabla _{\mathbf{R}}^2\left| {{\psi _\beta }} \right\rangle $ are derivative couplings between electronic states $\left| {{\psi _{\alpha, \beta} }} \right\rangle $ (or more precisely, polaritonic states here). The inner product amounts to integrating over electronic (and photonic) coordinate, but not nuclear one (see [S8,S9] for more details). Without light-matter coupling, the eigenstate of the electron-field Hamiltonian (eq.5 of the main text) the tensor product of the electronic state and photon fock state. The bare ground state $\left| {{\phi _0}\left( {\mathbf{R}} \right)} \right\rangle \left| {\text{0}} \right\rangle $, with the presence of interaction, becomes $\left| {{\psi _{0,0}}\left( {\mathbf{R}} \right)} \right\rangle $ (The two subscript $0$ denote electronic ground state and zero photon number).
	\begin{equation}
		\begin{aligned}
			\left| {{\psi _{0,0}}\left( {\mathbf{R}} \right)} \right\rangle  =& \left| {{\phi _0}\left( {\mathbf{R}} \right)} \right\rangle \left| {\text{0}} \right\rangle  + \sum\limits_{i,I} {\frac{{\left\langle {{\phi _i}\left( {\mathbf{R}} \right)} \right|\left\langle I \right|{{\hat H}_{{\text{el - em}}}}\left| {{\phi _0}\left( {\mathbf{R}} \right)} \right\rangle \left| {\text{0}} \right\rangle }}{{ - {E_{i0}}\left( {\mathbf{R}} \right) - I\Omega }}\left| {{\phi _i}\left( {\mathbf{R}} \right)} \right\rangle \left| I \right\rangle }  \hfill \\
			=& \left| {{\phi _0}\left( {\mathbf{R}} \right)} \right\rangle \left| {\text{0}} \right\rangle   + \sum\limits_i {\frac{{\left\langle {{\phi _i}\left( {\mathbf{R}} \right)} \right|{\mathbf{\hat d}}\left| {{\phi _0}\left( {\mathbf{R}} \right)} \right\rangle  \cdot {{\mathbf{E}}_{10}}}}{{{E_{i0}}\left( {\mathbf{R}} \right) + \Omega }}\left| {{\phi _i}\left( {\mathbf{R}} \right)} \right\rangle \left| 1 \right\rangle }   \hfill \\ 
		\end{aligned} 
	\end{equation}
	Only ${\left| {{\phi _i}\left( {\mathbf{R}} \right)} \right\rangle }$ and $\frac{{\left\langle {{\phi _i}\left( {\mathbf{R}} \right)} \right|{\mathbf{\hat d}}\left| {{\phi _0}\left( {\mathbf{R}} \right)} \right\rangle }}{{{E_{i0}}\left( {\mathbf{R}} \right) + \Omega }}$ depend on $\mathbf{R}$. Without resonance, i.e., no singularity in the denominator, these quantities depending on electronic wave function and energy typically vary with the inverse of Bohr radius $\frac{1}{a_0} \sim \sqrt {{E_{el}}{m_e}} $. Here $E_{el}$ is the typical electronic energy which is of the order of Rydberg energy or the gap between PESs, and $m_e$ is the electron mass. We can estimate ${{\mathbf{F}}_{\alpha ;00}}$ and ${G_{\alpha ;00}}$ as $a_0^{ - 1}$ and $a_0^{ - 2}$, which is the same as that in the standard analysis of Born and Oppenheimer's. Therefore, as in their result, the order of magnitude of the two terms $\frac{{{{\mathbf{F}}_{\alpha \beta }} \cdot {\mathbf{P}}}}{M}$, $\frac{{{G_{\alpha \beta }}}}{{2M}}$ are also ${E_{el}}{\left( {\frac{{{m_e}}}{M}} \right)^{3/4}}$ and ${E_{el}}\frac{{{m_e}}}{M}$ and can be neglected. 
	%----We can also write the follows which may be more tedious: ----
	%We can estimate ${{\mathbf{F}}_{\alpha ;00}}$ and ${G_{\alpha ;00}}$ as $a_0^{ - 1}$ and $a_0^{ - 2}$. The momentum $\mathbf{P}$ is roughly $\sqrt {{E_{{\text{vib}}}}M} $ where the vibrational energy ${E_{{\text{vib}}}} \sim E_{el} (\frac{m_e}{M})^{\frac{1}{2}}$, according to the result of Born and Oppenheimer's\cite{born1985quantentheorie}. With the above knowledge, the order of magnitude of the two terms $\frac{{{{\mathbf{F}}_{\alpha \beta }} \cdot {\mathbf{P}}}}{M}$, $\frac{{{G_{\alpha \beta }}}}{{2M}}$ are ${E_{el}}{\left( {\frac{{{m_e}}}{M}} \right)^{3/4}}$ and ${E_{el}}\frac{{{m_e}}}{M}$.
	
	In the main text, we utilized the Arrhenius equation to evaluate the rate constant. To improve the accuracy, one needs to take the oscillations of nucleus into account. It is worth noting that even at absolute zero temperature, nucleus still exhibit oscillatory motion due to zero-point fluctuations. In the main text, we took the ``activation energy" $E_a$ in the exponent $\exp \left( { - {E_a}/{k_B}T} \right)$ as the classical barrier height $E_\text{barrier}$ on the PES. A superior theory for predicting the rate constant from the information of PESs is the transition state theory (TST), where the rate is given by [S10]
	\begin{equation*}
		{k_\text{TST}} = \frac{{{k_B}T}}{h}\frac{{{Z_{{\text{\ddag}}}}}}{{{Z_{{\text{rea}}}}}}{e^{ - \frac{{{E_a}}}{{{k_B}T}}}}.
	\end{equation*}
	Here, ${k_B}$ and $h$ are the Boltzmann and Planck constants, the $Z_\ddag$ is the partition function of the transition state without the contribution from the reactive coordinate (A transition state is the nuclear configuration with the highest energy along the reaction coordinate.) and $Z_\text{rea}$ is the partition function of the reactant state. The activation energy is ${E_a} = {E_{{\text{barrier}}}} + \frac{1}{2}\sum\limits_i {\hbar {\omega _{\ddag ,i}}}  - \frac{1}{2}\sum\limits_i {\hbar {\omega _{{\text{rea}},i}}} $. ${E_{{\text{barrier}}}}$ is the barrier height on the PES, and the latter two terms are two set of zero-point energy obtained by treating the transition state ($\ddag$) and the reactant state (rea) as multi-dimensional harmonic oscillators. Note that the zero-point energy of the transition state along the reaction coordinate is zero since the motion along this direction is unbounded. We ignore the factors other than the exponential factor since they do not differ a lot between the enantiomers. In our model of the hydrogen-missing helicene molecule, taking into account only the reaction coordinate, the activation energy is 
	\begin{equation*}
		{E_a} = {E_{{\text{barrier}}}} - \frac{1}{2}\hbar {\omega _\nu },
	\end{equation*}
	where $\omega_\nu$ is the vibrational frequency of the reaction coordinate around the reactant region. An illustration is shown in Fig.S1. 
 	\begin{figure}[!htb]
		\centering
		\includegraphics[width=16cm, angle=0]{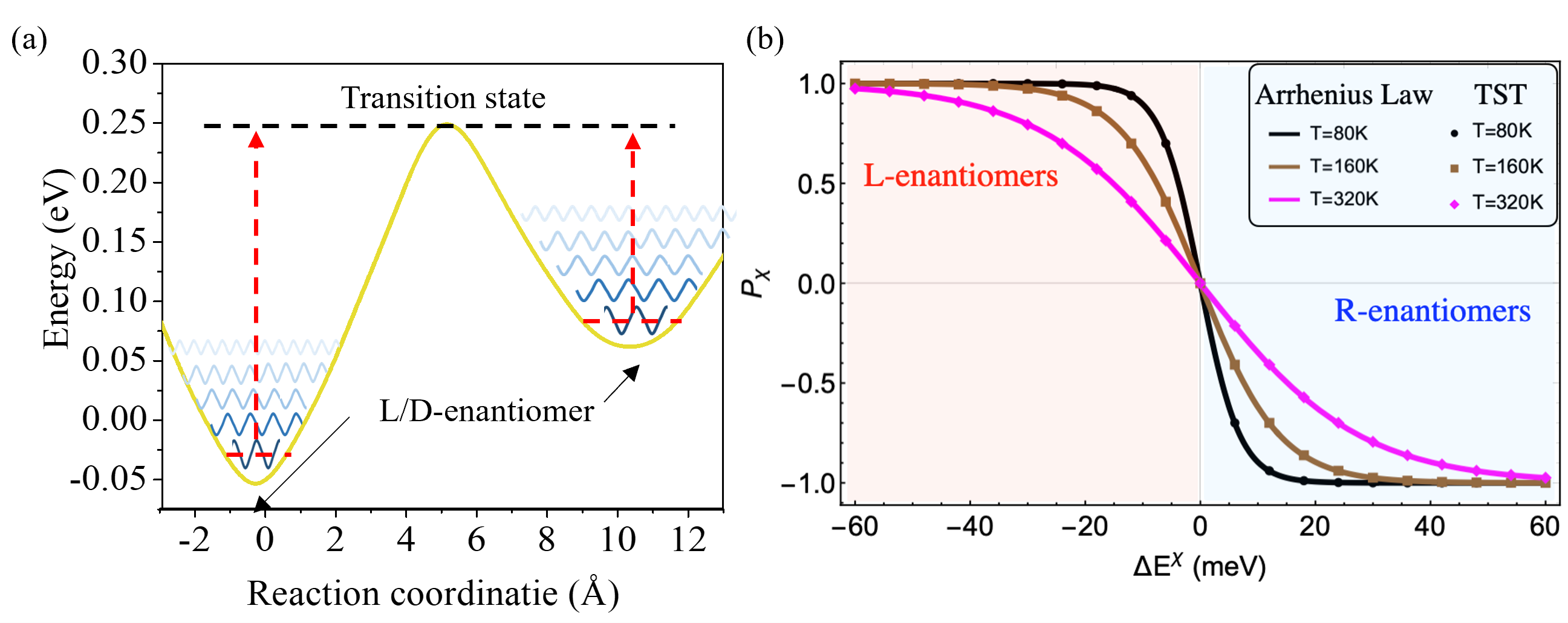}
		\caption{{ (a) Schematically illustration of the activation energy. The yellow line is the PES with the energy shift induced by cavity, and the wavy lines depict the vibrational levels around two (sub-)stable states (not the exact positions of the energy levels). It is important to note that the activation energy is not solely the energy difference between the transition state and the reactant state, but rather reduced by the vibrational zero-point energy. 
        (b) The chirality-selective rate is plotted as a function of the chiral energy shift at various temperatures, considering both the simple Arrhenius law and transition state theory (TST). The solid lines represent cases where the activation energy is directly given by the energy shift $\Delta E^\chi$ (same in the Fig.4 in the main text). In contrast, the dots represent cases where the activation energy is determined according to TST by subtracting the zero-point energy, which is a function that depends on $\Delta E^\chi$.}
     }
		\label{fig_5}
	\end{figure}

 In additional to the energy change of the two enantiomers induced by chiral energy shift, there is also a modification in the vibrational zero-point energy. In the vicinity of each local minimum, the potential energy surface (PES) can be approximated as a quadratic function of the deviation from the minimum, denoted as $\Delta R$. The bare molecule PES is thus ${E_0} + \frac{1}{2}M\omega _\nu ^2{\left( {\Delta R} \right)^2}$ with the effective mass of this coordinate, $M$, which can be estimated as one carbon atom mass; focusing on the opposite energy shift of the two enantiomers, Casimir-Polder effect modifies the PES to ${E_0} \pm \Delta E_0^\chi  + \frac{1}{2}\left( {M\omega _\nu ^2 \pm b} \right){\left( {\Delta R} \right)^2}$ around the two enantiomer states, respectively. Here $\Delta E_0^\chi $ is the chiral energy shift studied in the main text, and $b$ is the change of the coefficient of the quadratic term that can be obtained by fitting the numerical calculated PES. Calculating the shift of $\omega_\nu$, $\Delta \omega_\nu$, by $M\omega _\nu ^2 \pm b \triangleq M{\left( {{\omega _\nu } \pm \Delta {\omega _\nu }} \right)^2}$, we have found that for a energy shift $\Delta E_0^\chi  \approx 53{\text{meV}}$ (with 100 molecules. See Fig.3 in the main text.), the correction from zero-point energy is smaller by two orders of magnitude, $\frac{1}{2}\hbar \Delta {\omega _\nu } \approx  - 0.2{\text{meV}}$, which can be safely ignored.
The Fig.\ref{fig_5}(b) shows the influence of nuclear oscillations to the chirality-selective rate (Eq.(17)). It shows that the Arrhenius equation provides a highly accurate approximation to the TST.
	
 The modification from the zero-point energy is somewhat ``quantum" in nature since it takes into account the quantum characteristics of the confined degree of freedom. However, the transition state theory (TST) as a whole is typically regarded as a classical or semi-classical theory, incapable of replacing a complete quantum dynamic calculation of the reaction rate. Nonetheless, TST has been generally proved to be a reliable approximation, exhibiting a direct link between the rate and the energy barrier, which has been employed in this work.\\

\section{Detail methods of density functional theory calculations}
The density functional theory calculation was performed with norm conserving pseudopotential on the basis set of projector augmented plane waves. A cutoff of 400 eV was applied to the plane waves. PBE functional was used to deal with the electron-electron exchange and correlation interaction. To show the spin polarization, spin orbital coupling (SOC) was turned off when the energy level was calculated. SOC was turned on for all the other calculations. A vacuum space larger than 10 Å was created in all three directions to decouple the periodic imagines. All atoms were relaxed until the force on each atom is smaller than 0.01 eV/Å.  The method of nudged elastic band (NEB) was applied to search the transition states. { The reaction coordinate is defined as $ R_j=\sqrt{\displaystyle\sum\limits_{m=1}^{j} \sum_i(\bf{R}_{i,m}-\bf{R}_{i,m-1})^2}$, where {\bf{R}} is the atomic position vector, \textit{j} indexes the NEB step, and \textit{i} is the atomic index. Such a concept is generally used in chemistry.[S11-12] The reaction coordinate here represents the average atomic displacement from a left-handed molecule to a right-handed molecule. It is a one-dimensional abstract coordinate showing the progress along chiral reaction route.}

\begin{center}
\textbf{Supplemental References}
\end{center}

[S1] 
A. Abrikosov, L. Gorkov, I. Dzyaloshinski, and R. Silverman, Methods of Quantum Field Theory in Statistical Physics, Dover Books on Physics (Dover Publications, 2012), ISBN 9780486140155.

[S2]
S. Y. Buhmann, Dispersion Forces I: Macroscopic quantum electrodynamics and ground-state Casimir, Casimir–Polder and van der Waals forces, vol. {\bf 247} (Springer, 2013).

[S3] 
Q.-D. Jiang and F. Wilczek, Physical Review B {\bf 99}, 165402 (2019).

[S4]
Notice that in this gauge the electromagnetic field operator is $\hat {\mathbf{B}} = \nabla  \times \hat {\mathbf{A}},\;\hat {\mathbf{E}} =  - i\left[ {\hat H,\hat {\mathbf{A}}} \right]$, and thus the relations between these "Green's function"'s.

[S5]
Besides, we mention that the free space contribution to Green's function, or homogeneous Green's function, is symmetric $G_{ij}^{free}\left( {{\mathbf{r}},{\mathbf{r}}'} \right) = G_{ji}^{free}\left( {{\mathbf{r}},{\mathbf{r}}'} \right)$, so ${\text{Tr}}\nabla  \times {\mathbf{G}}\left( {{{\mathbf{r}}_M},{{\mathbf{r}}_M}} \right)$ must vanish.

[S6]
J. Galego, F. J. Garcia-Vidal, and J. Feist, Physical Review X {\bf 5}, 041022 (2015).

[S7]
M. Kowalewski, K. Bennett, and S. Mukamel, The Journal of Chemical Physics {\bf 144}, 054309 (2016).

[S8]
M. Born and R. Oppenheimer, Annalen der Physik {\bf 389}, 457 (1927).

[S9]
J. C. Tully, Theoretical Chemistry Accounts {\bf 103}, 173 (2000).

[S10]
N. E. Henriksen and F. Y. Hansen, Theories of molecular reaction dynamics: the microscopic foundation of chemical kinetics (Oxford University Press, 2018).

[S11] G. Henkelman and Hannes J{\'o}nsson,  The Journal of chemical physics {\bf 113}, 9978 (2000).

[S12]
G. Henkelman, B. P. Uberuaga and H. J{\'o}nsson, The Journal of chemical physics, {\bf 113}, pp.9901 (2000).

\end{widetext}

\end{document}